\documentclass[prb,twocolumn,aps,showpacs,preprintnumbers,amsmath,amssymb,citeautoscript]{revtex4}

\usepackage[dvips]{graphicx}% Include figure files
\newcommand{\bra}[1]{\left<#1\right|}
\newcommand{\ket}[1]{\left|#1\right>}

\newcommand{\ba}{{\bf a}}

\newcommand{\br}{{\bf r}}
\newcommand{\bk}{{\bf k}}

\newcommand{\be}{{\bf e}}
\newcommand{\bm}{{\bf m}}
\newcommand{\bn}{{\bf n}}

\newcommand{\bA}{{\bf A}}
\newcommand{\bB}{{\bf B}}

\newcommand{\cH}{{\cal H}}

\newcommand{\f}{\frac}

\newcommand{\df}{\nabla}
\newcommand{\db}{\overline\nabla}

\newcommand{\Eq}[1]{Eq.~(\ref{#1})}
\newcommand{\Fig}[1]{Fig.~\ref{#1}}
\newcommand{\Ref}[1]{(\ref{#1})}

\newcommand{\av}[1]{\left<{#1}\right>}

\begin{document}

%\preprint{DRAFT!!!  NOT FOR DISTRIBUTION!}

\title{Topological order and the deconfinement transition in the (2+1)
  dimensional compact Abelian Higgs model}

\author{Anders Vestergren}
\email{anders@condmat.physics.kth.se}

\author{Jack Lidmar}
\email{jlidmar@kth.se}

\affiliation{
Department of Physics,
Royal Institute of Technology,
AlbaNova,
SE-106 91 Stockholm,
Sweden
}

\date{\today}

\begin{abstract}
 We study an Abelian compact gauge theory minimally coupled to bosonic
 matter with charge $q$, which may undergo a
 confinement--deconfinement transition in (2+1)D.  The transition is
 analyzed using a nonlocal order parameter $\tilde W$, which is
 related to large Wilson loops for fractional charges.  We map the
 model to a dual representation with no gauge field but only a global
 $q$-state clock symmetry and show that $\tilde W$ correspond to the
 domain wall energy of that model.  $\tilde W$ is also directly
 connected to the concept of topological order.  We exploit these
 facts in Monte Carlo simulations to study the detailed nature of the
 deconfinement transition.
\end{abstract}

\pacs{
74.20.-z, % (Theories and models of superconducting state)
11.15.Ha, % (Lattice gauge theory)
71.10.Hf % (Non-Fermi-liquid ground states, electron phase diagrams
	 % and phase transitions in model systems) 
}

%\keywords{} % Use showkeys class option if keyword display desired

\maketitle

\section{Introduction}

The appearance of excitations with fractional quantum numbers in
strongly correlated systems has received much interest in recent
years~\cite{baskaran:580, ioffe:8988, nagaosa:966, lee:5621,
nagaosa-lee, Read91, Sachdev91, sachdev03, wen:503, senthil-fisher}.
The splintering of the electron into a neutral spinon and a charged
spinless holon, and the resulting spin-charge separation, was proposed
early after the discovery of the superconducting high-$T_c$ cuprates
as a route to explain the non-Fermi liquid behavior observed in these
systems.  The fractionally charged Laughlin quasi-particles in the
fractional quantum hall effect is another famous example.  While
spin-charge separation is well established in one dimensional electron
gases, the occurrence in higher dimensions remains poorly understood.
Usually, the fractionalization in dimensions larger than one, is
imagined to occur via a confinement-deconfinement transition of an
effective low energy gauge theory description of the
system~\cite{mudry-fradkin,mudry-fradkin-2}.  A number of such gauge
theories have been proposed for the underdoped cuprates, including
$U(1)$~\cite{baskaran:580, ioffe:8988, nagaosa:966, lee:5621,
nagaosa-lee, Read91, Sachdev91, sachdev03, DHLee00,Ichinose2001},
$SU(2)$~\cite{ wen:503}, and $Z_2$~\cite{ senthil-fisher}-gauge
theories.  In addition, the possibility of fractionalized phases in
models containing only bosons has been proposed~\cite{senthil-montro}.
The fractionalized phases are not ordered in a conventional sense,
which breaks any obvious symmetries of the system.  Instead they are
characterized by a special kind of order called topological
order~\cite{wen91}.  Topological order is robust against local
perturbations and furthermore leads to a ground state degeneracy.  It
has been proposed that this topologically protected degeneracy could
serve as a building block for qubits in a quantum
computer~\cite{Feigleman-Ioffe,kitaev}.
An important prediction is that in finite sized systems the degeneracy
is lifted due to tunneling processes, leading to a splitting $\Delta
\sim e^{-L/\xi}$~\cite{wen91}.

In this paper we study the relation between topological order and
deconfinement in one of the simplest models displaying topological
order --- the compact Abelian Higgs model.  It should be noted that a
realistic theory of the high-$T_c$ cuprates would require also the
inclusion of gapless nodal quasiparticles, which are here
neglected~\cite{tesanovic,herbut,herbut-seradjeh,hermele,kleinert-nogu-sudbo}.
We thus consider a (2+1)D compact $U(1)$ gauge theory minimally
coupled to a charge-$q$ bosonic matter field, with an Euclidean action
\begin{equation}					\label{eq:action}
S = - J \sum_{\br\mu} \cos(\df_\mu\theta_\br - q A_{\br\mu})
  - \f{1}{g} \sum_{\br\mu} \cos( B_{\br\mu} ),
\end{equation}
where $\theta_\br$ and $A_{\br\mu}$ are compact phases $\in [0,2\pi)$
living on the sites and links of a 3D simple cubic lattice,
respectively.  $B_{\br\mu} = \epsilon_{\mu\nu\lambda}\df_\nu
A_{\br\lambda}$ is the dual field strength, with the lattice difference
operators defined by $\df_\mu f_\br = \db_\mu f_{\br + \be_\mu} = f_{\br +
\be_\mu} - f_\br$.
The pure gauge theory in absence of matter is always confining in
(2+1)D, due to the proliferation of instantons~\cite{poly77}.  (In
(3+1)D a deconfined Coulomb phase is also possible.~\cite{poly77})
Partly because of this much attention has focused on $Z_2$-gauge
theories~\cite{senthil-fisher}, which do allow a
confinement-deconfinement transition in (2+1)D.  Alternatively, when
the $U(1)$ model is coupled to a matter field which does not belong to
the fundamental representation (i.e., when the charge $q$ of the boson
differs from unity in \Eq{eq:action}), a phase transition is
possible~\cite{frad79}.  This could happen, e.g., if the bosonic field
describes the pairing of
spinons~\cite{nagaosa-lee,wen-spinon-pairing}.  Indeed, the $U(1)$
gauge theory coupled to a charge 2 boson is equivalent to a
$Z_2$-gauge theory in the limit $J \to \infty$.  In such a system, the
fractionalized phase would then correspond to a phase where the
spinon-pairs breaks up into separate free excitations.
More generally, the model with gauge charge $q$ reduces to a
$Z_q$-gauge theory in the limit $J \to \infty$.  
In the opposite limit $g \to 0$, the gauge field fluctuations freeze
out and a 3D $XY$ model (with a global U(1) symmetry) results.
One may ask how the transition interpolates between these two extremes
with increasing screening length $\lambda = 1/\sqrt{Jgq^2}$.  The
naive expectation would be that the $\lambda\to\infty$ transition
would represent an unstable fixed point and that any finite value of
$\lambda$ would flow to zero upon renormalization, giving a transition
in the $Z_q$ universality class.  In this case the 3D $XY$ point at
$\lambda\to\infty$ would be an isolated point in the phase diagram.

Recently, large scale Monte Carlo simulations of the (2+1)D compact
$U(1)$ Higgs model were performed by Sudb{\o} {\em et al.}\cite{sudbo}
and Smiseth {\em et al.} \cite{smiseth} for several values of the
charge $q>1$.  The results of these indicated that the phase diagram
might be surprisingly complicated, with continuously varying critical
exponents along the transition line between the $Z_q$ and 3D $XY$
values.  This would imply that the transition line for intermediate
$\lambda$ might be described by a line of fixed points.
These interesting results were obtained from a finite size scaling
analysis of the third moment of the action, which is a purely
thermodynamic observable.  Although their method seems to work very
well, it would be desirable to have a clear order parameter which
distinguishes the confining and deconfining phases, especially in
light of their surprising results.

%%%%%%%%%%%%%%%%%%% {Outline paper}

In this paper we employ a nonlocal order parameter $\tilde W$,
previously introduced in Ref.~\onlinecite{ves1}, which provides a
direct probe of the confining properties of the theory.  We show that
this order parameter can be mapped via a duality transformation to the
domain wall energy of a $Z_q$ clock model, and present an in depth
discussion of its relation to topological order, including many of the
details left out in Ref.~\onlinecite{ves1}.  We obtain the finite size
scaling properties of $\tilde W$, discuss its relation to the
splitting $\Delta$ of the nearly degenerate ground states, and the
energy gap to vortex excitations in the deconfined phase.  Finally, we
use it in Monte Carlo simulations to study the detailed nature of the
deconfinement transition of the model, and compare with the earlier
results mentioned above.

The paper is organized as follows.  In Sec.~\ref{sec:wilson-loops} we
reformulate the theory in terms of its topological defects, which are
vortices and instantons, and discuss how $\tilde W$ can distinguish
the different phases.  Going over to another dual formulation a model
with a global $q$-state clock symmetry (and no gauge fields) obtains,
and $\tilde W$ translates into the domain wall energy of that model,
or equivalently the energy cost for twisting the boundary conditions
by $2\pi/q$, see Sec.~\ref{sec:mapping}-\ref{sec:domain-wall}.  In
Sec.~\ref{sec:topo}, we discuss the relation to topological order,
adding details and generalizations of the ideas discussed in Ref.\
\onlinecite{ves1}.  In Sec.~\ref{sec:MC} we describe Monte Carlo
simulations of the (2+1)D compact Higgs model carried out in the
loop-gas representation of flux lines and monopoles.  Our results and
conclusions are discussed in Sec.~\ref{sec:results} and
\ref{sec:summary}.

\section{Duality transformations, Wilson loops and topological order}

In order to transform \Eq{eq:action} to alternative representations it
is convenient to instead study the Villain version of the model,
which allow the action to be transformed to dual representations
exactly, without further approximations, while retaining the important
physics.
In the Villain approximation the action is
\begin{equation}					\label{eq:Villain}
  S =  \sum_{\br\mu}
  \f{J}{2} \left( \df_\mu\theta_\br + 2\pi n_{\br\mu} - q A_{\br\mu}
  \right)^2 + \sum_{\br\mu} \f{1}{2g} \left( B_{\br\mu} + 2\pi k_{\br\mu}
  \right)^2,
\end{equation}
where $n_{\br\mu}$ and $k_{\br\mu}$ are dummy integers to be summed
over.  The summation over $k_{\br\mu}$ has the effect of making the
energy cost of putting an integer multiple of $q$ flux quanta $\Phi_0
= 2\pi/q$ through a plaquette of the lattice vanish.  The net effect
of this is the proliferation of Dirac strings, which leads to the
appearance of magnetic monopoles of charge $q \Phi_0$ ($=2\pi$).

A standard set of transformations~\cite{frad79,ein78} allows
\Eq{eq:Villain} to be rewritten in terms of flux lines and monopoles,
interacting via the action
\begin{equation}				\label{eq:vortexlinerepr}
  S = \sum_{\br\br'} \f{K}{2} \bm_\br \cdot V_{\br\br'} \bm_{\br'}
    + \f{K\lambda^2 q^2}{2} N_\br V_{\br\br'} N_{\br'}.
\end{equation}
Here $\bm_\br \in \mathbb{Z}^3$ is the vorticity on the links and
$N_\br \in \mathbb{Z}$ is the monopole charge,\footnote{Strictly
speaking the flux lines and monopoles live on the dual lattice of the
original model.  Since this is also a cubic lattice, shifted by half a
lattice constant in the (111) direction we use the same index $\br$ to
denote sites on this dual lattice.} and $K = (2\pi)^2 J$.  The flux
lines are constrained to start or end on the monopoles only in quanta
of $q$, i.e., $\nabla \cdot \bm_\br = qN_\br$.
The flux lines and monopoles interact through a screened coulomb
interaction, defined by its Fourier transform $V^{-1}_\bk =
\kappa_\bk^2 + \lambda^{-2}$, where $\kappa^2 = \sum_\nu
\kappa_{\bk\nu}^2$, and $\kappa_{\bk\nu} = 2\sin(k_\nu /2)$, i.e.,
\begin{equation}				\label{eq:interaction}
  V_{\br} = \f{1}{L^3} \sum_\bk
  \f{e^{i \bk \cdot \br}}{ \kappa_\bk^2 + \lambda^{-2} },
\end{equation}
which for large distances is a screened coulomb interaction $V(r) =
e^{-r/\lambda}/4\pi r$ with a screening length $\lambda$ given by
$\lambda^{-2} = K g / \Phi_0^2 = J g q^2$.  In the limit $\lambda \to
0$ the interaction reduces to an onsite interaction $V_{\br} =
\lambda^2 \delta_{\br}$, and the action becomes
\begin{equation}
   S = \f{\Phi_0^2}{2g} \sum_\br \bm_\br^2.
\end{equation}

\begin{figure*}
{\includegraphics[width=0.45\linewidth]{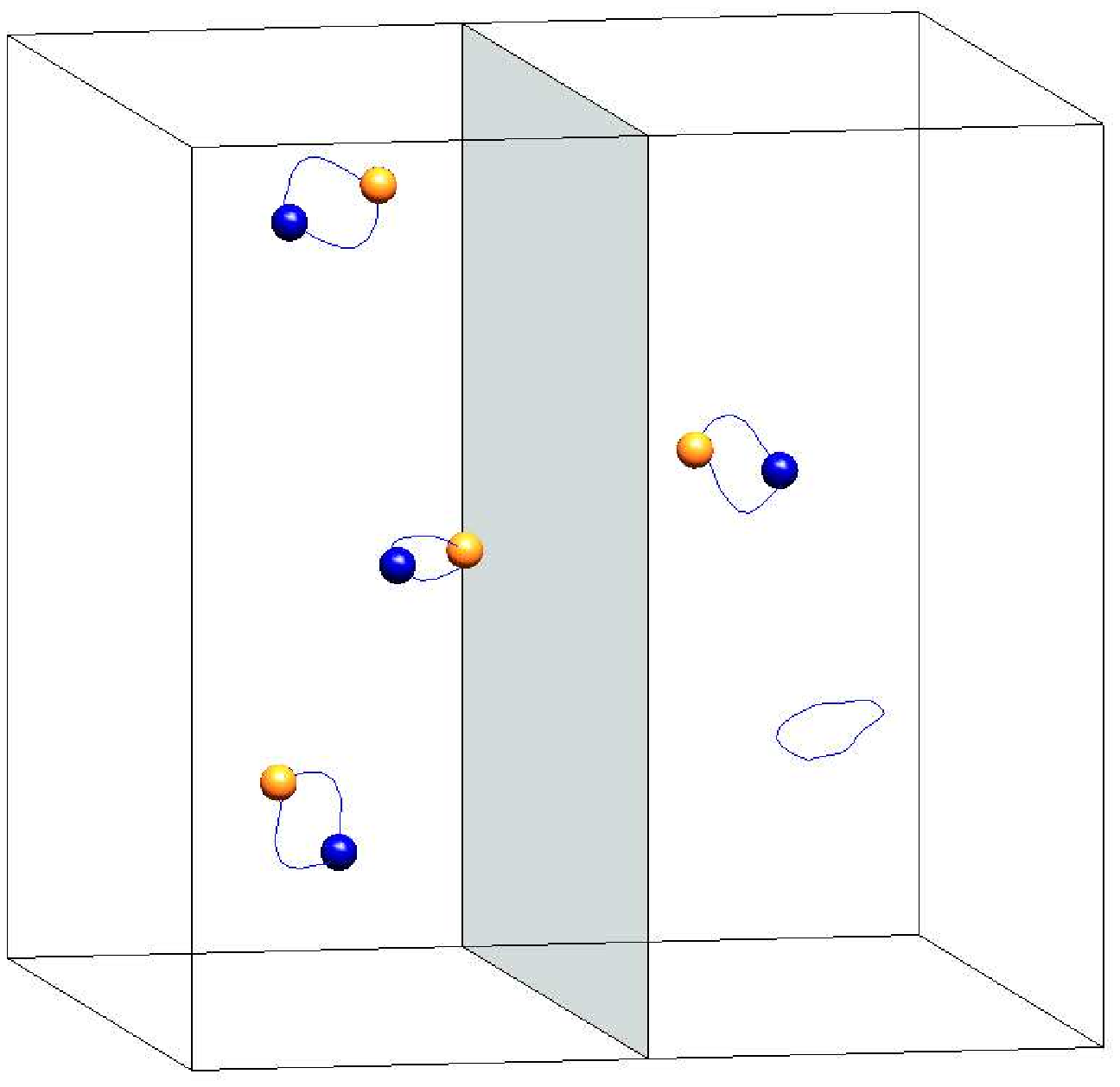}
\includegraphics[width=0.45\linewidth]{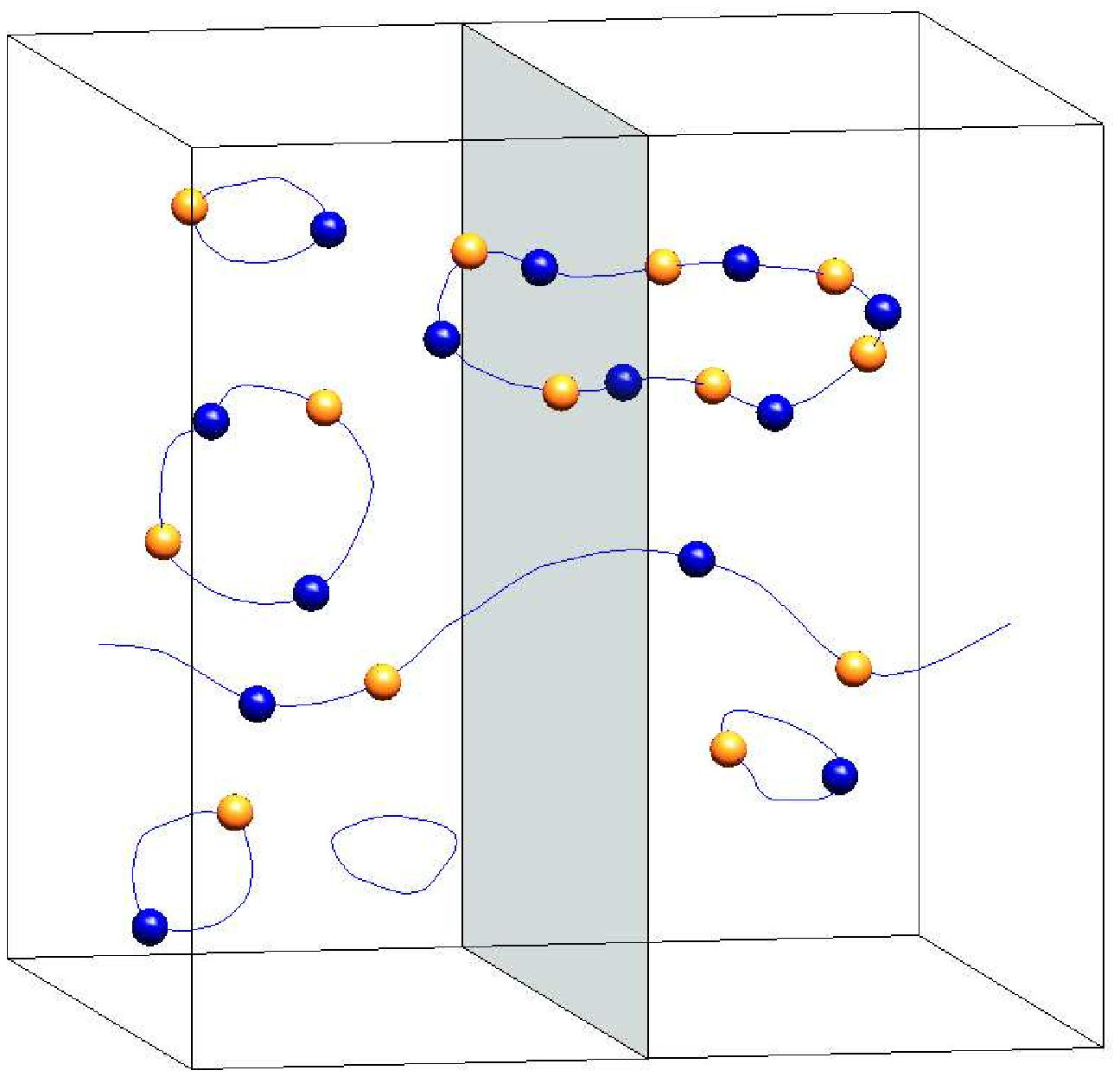}}
\caption{						\label{fig:wilson}
(Color online)
Possible phases of the system.  To the left the topological ordered,
deconfined phase, where vortices cost much energy.  To the right the
percolating vortex tangle in the confined phase.  The large Wilson
loop $\tilde W$ surrounds the shaded area as indicated in the figures.
}
\end{figure*}

The constraint $\nabla \cdot \bm = qN$ means that a $+$($-$) monopole
has precisely $q$ outgoing (incoming) flux lines.  For $q>1$ one may
then envisage two distinct phases of the system.  Either the monopoles
mostly pair up in neutral pairs, where the $q$ outgoing flux lines of
the plus charge all end at the minus charge of the pair, as
illustrated in \Fig{fig:wilson}(a).  This phase will be realized in
the limit of large $K$, where the flux lines cost a lot of energy.
As discussed in the next section, this magnetically confining phase
corresponds to deconfined fractional electric charges.
The other possibility is that the outgoing flux lines mostly end on
different monopoles, forming a large connected tangle of flux lines
and monopoles, which percolate through the whole system, as in
\Fig{fig:wilson}(b).
Recent simulations~\cite{chernodub} support this interpretation.
A transition between these two geometrically different phases is
expected at some critical value of the coupling $K_c$ (or $g_c$).
This picture can be made more precise by studying the Wilson loop for
fractional charges.

\subsection{Wilson loops}			\label{sec:wilson-loops}

The presence of percolating flux lines in the system can be detected
by counting the number of flux lines crossing a given surface
$\mathcal{S}$ that form a cross-section of the system, see
\Fig{fig:wilson}.  We assume here periodic boundary conditions for the
vorticity $\bm$ in all directions, i.e., the system has the topology
of a 3-torus.
This corresponds to allowing twists in the boundary conditions for
$\bA$ in \Eq{eq:action} or \Ref{eq:Villain}~\cite{ves1}.
Let the \emph{winding number} $M_\nu$ be the number of
such flux line crossings (counted with signs) for a cross-section
perpendicular to the $\nu$-direction, i.e.,
\begin{equation}					\label{eq:winding}
  M_\nu = \int_\mathcal{S} \bm \cdot d\mathbf{S}.
\end{equation}
(For notational convenience we use a continuum formulation here.)  The
precise location of the surface does not matter.  In the paired phase
this number will be an integer multiple of $q$, whereas in the
percolating phase it can be any integer.  It turns out that precisely
this property is captured by a large Wilson loop for fractionally charged
test particles,
\begin{equation}
\tilde{W} \equiv {W}(\mathcal{C}) = \av{\exp\left({i \oint_\mathcal{C}
    \bA \cdot d\br}\right)},
\end{equation}
where the loop $\mathcal{C}$ encircles the surface $\mathcal{S}$.
The expression in the exponent,
\begin{equation}
  \oint_\mathcal{C} \bA \cdot d\br = 
  \int_\mathcal{S} \bB \cdot d\mathbf{S} = 
  \f{2\pi}{q} \int_\mathcal{S} \bm \cdot d\mathbf{S},
\end{equation}
equals $2\pi/q$ times the winding number $M_\nu$, and the Wilson loop
for a loop with this special geometry is thus given by
\begin{equation}						\label{eq:W}
  \tilde{W} = \av{\exp\left( \f{2\pi i}{q} M_\nu \right)}.
\end{equation}
This becomes one in the paired phase where $M_\nu$ is an integer
multiple of $q$ [\Fig{fig:wilson}(a)].  In the percolating phase
[\Fig{fig:wilson}(b)] no particular value of the winding number
$M_\nu$ will be favored, making the $\tilde W$ approach zero with
increasing system size.  Note that it is the global topological
structure of the flux line tangle that determines the value of $\tilde
W$, while small scale entanglement is unimportant.

A Wilson loop decaying with an area law, ${W}(\mathcal{C}) \sim
\exp(-c L^2)$ implies a linearly confining attraction between static
fractional charges, while a perimeter law, ${W}(\mathcal{C}) \sim
\exp(-c L)$, means that such charges would be deconfined.
Note that the behavior of the charges is opposite of the monopoles:
The charges are deconfined when the monopoles form pairs and confined
when the pairs break up.
Remember that the charge of the bosons in the original model is $q$,
so the Wilson loop for charge-$q$ test particles would include a
factor $q$ in the exponent, making $\tilde W$ equal to 1 exactly in
both phases.
For fractional charges, however, it changes from unity to an area law
$\tilde{W} \sim \exp( -c L^2)$, when going through the
transition (deep in the confined phase one finds $c = g/2$).
Another way to say this is that the charge-$q$ dynamic matter cannot
screen out the fractionally charged test particles.

The absence of a perimeter law for $\tilde W$ in the deconfined phase
is a consequence of the special geometry of a loop which covers a
whole cross-section of the system, and in some sense has no perimeter
due to the periodic boundary conditions.  This makes our order
parameter $\tilde W$ different from ordinary Wilson loops, and is a
big advantage since it makes it easier to analyze its scaling behavior
as discussed in Sec.~\ref{sec:domain-wall}.  For a finite sized loop
in a much larger system the Wilson loop would follow a perimeter law
rather than approach unity in the deconfined phase.  This is due to
configurations where part of the flux connecting a
monopole-anti-monopole pair would pass through the loop and part
outside the loop.  In the paired phase this can happen only very close
to the loop itself, hence giving rise to a perimeter law.
In the next two sections we describe the relation of the Wilson loop
order parameter $\tilde W$ to the domain wall energy in a dual
clock-symmetric model.

\subsection{Mapping to a model with clock symmetry}	\label{sec:mapping}

% Clock model

We will now map \Eq{eq:vortexlinerepr} to a dual model with a
$q$-state clock symmetry.  The idea is to start from the action
\begin{eqnarray}				\label{eq:vortexlinerepr2}
  S \nonumber &=& \sum_{\br\br'} \f{K}{2} \bm_\br \cdot V_{\br\br'} \bm_{\br'}
    + \f{K\lambda^2 q^2}{2} N_\br V_{\br\br'} N_{\br'} \\
    &+& \sum_\br i \chi_\br \left( \nabla \cdot \bm_\br - qN_\br \right),
\end{eqnarray}
where the constraint $\nabla \cdot \bm = q N$ is implemented via a
Lagrange multiplier $\chi_\br$, and then integrate out the flux lines
and monopoles to get an effective action for $\chi$.  We decouple the
vortex-vortex and monopole-monopole interactions with the help of two
auxiliary fields, $\ba$ and $\phi$, to get
\begin{eqnarray}
S \nonumber &=& \sum_{\br\br'} \f{1}{2K} \ba_\br \cdot V^{-1}_{\br\br'} \ba_{\br'}
    + \f{1}{2K\lambda^2} \phi_\br V^{-1}_{\br\br'} \phi_{\br'} \\
    &+& \sum_\br i \ba_\br \cdot \bm_\br + i \phi_\br qN_\br
    + i \chi_\br \left( \nabla\cdot \bm_\br - qN_\br \right) 
\end{eqnarray}
Now the summation over the integers $\bm_\br$ and $N_\br$ can be
performed, leading to
\begin{equation}
  \ba = \nabla\chi + 2\pi\bn, \quad \phi = \chi + 2\pi k / q,
  \qquad \bn, \, k \; \mbox{integers}.
\end{equation}
We then obtain the action
\begin{eqnarray}				\label{eq:clock-model}
  S \nonumber &=& \sum_{\br\br'} \f{1}{2K} \left( \nabla\chi_\br + 2\pi\bn_\br \right)
               V^{-1}_{\br\br'}\left( \nabla\chi_{\br'} + 2\pi\bn_{\br'} \right)\\
    &+& \f{1}{2K\lambda^2} \left(\chi_\br + 2\pi k_\br/q \right)
  V^{-1}_{\br\br'}\left( \chi_{\br'} + 2\pi k_{\br'}/q \right).
\end{eqnarray}
As announced this model has only a global $q$-state clock symmetry and
contains no gauge fields and only short range interactions.
Rewriting it in Fourier space gives
\begin{eqnarray}				\label{eq:clock-model-f}
  S &=& \f{1}{2K\Omega} \sum_\bk \left( \kappa^2 + \lambda^{-2}\right)
      \left| \boldsymbol{\kappa}\chi_\bk + 2\pi\bn_\bk \right|^2
      \nonumber \\
    &+& \f{1}{\lambda^2} \left( \kappa^2 + \lambda^{-2}\right)
      \left| \chi_\bk + 2\pi k_\bk/q \right|^2.
\end{eqnarray}
Obviously $V_\bk^{-1} = \kappa^2 + \lambda^{-2}$ is short ranged and
may at least for small values of $\lambda$ (i.e.\ large $g$) be
replaced by a completely local interaction, $V_{\br\br'}^{-1} \approx
\lambda^{-2}\delta_{\br\br'}$.  The corrections enter with higher
order derivative and should therefore be irrelevant at small
$\lambda$.
Dropping these and ``undoing'' the Villain approximation now gives
\begin{equation}
  S \approx - \f{1}{K\lambda^2} \left\{ \sum_{\br\mu}
  \cos(\df_\mu \chi_\br)
  + \f{1}{q^2 \lambda^2} \sum_\br \cos(q\chi_\br)\right\}
\end{equation}
which clearly has a $q$-state clock symmetry.  For small $\lambda^2$
it is legitimate to replace the continuous variable $\chi_\br$ by a
discrete one $\chi_\br \approx 2\pi k_\br /q$ leading to
\begin{equation}				\label{eq:clock2}
  S \approx - \f{1}{K\lambda^2} \sum_{\br\mu}
  \cos(2\pi\df_\mu k_\br/q) ,
\end{equation}
which is just the familiar $q$-state clock model.
Equation~\Ref{eq:clock2} is exact in the limit $\lambda \to 0$, where
the well known duality between $Z_q$-gauge theory and the $q$-state
clock model is regained.  Equation~\Ref{eq:clock-model} remains valid
for arbitrary $\lambda$.

Note that for fundamental matter, when $q=1$, \Eq{eq:clock-model}
essentially describes an $XY$ model in a finite external magnetic
field, which does not have a phase transition; this agrees
with Ref.~ \onlinecite{nagaosa-lee}.

\subsection{Domain wall energy		\label{sec:domain-wall}}

We now proceed to show that the domain wall energy in the $q$-state
clock symmetric model, \Eq{eq:clock-model}, corresponds precisely to
the large Wilson loop $\tilde W$ in the original compact gauge theory.
Consider a finite sized system with side length $L$ and periodic
boundary conditions.  A domain wall can be forced into the system by
changing the boundary conditions such that $\chi(\br + L\be_\nu) =
\chi(\br) + \Delta\chi$, $\Delta\chi = 2\pi k/q$.  The domain wall
energy is defined as the free energy cost $\Delta F$ for introducing
such a twist.  It is convenient to make a change of variables in the
twisted system to get back to periodic boundary conditions by setting
$\chi(\br) = \theta(\br) + \Delta\theta(\br)$, where $\theta$ obeys
periodic boundary conditions, and $\Delta\theta(\br) = $ constant
except across a surface $\mathcal{S}$ occupying a cross section of the
system perpendicular to the direction $\nu$, where $\Delta\theta(\br)$
jumps discontinuously by $2\pi k/q$.  We can then evaluate the domain
wall energy as
\begin{equation}
  \Delta F = F_\mathrm{twisted} - F_\mathrm{untwisted}
  = - \ln\left< e^{-\Delta S} \right>
\end{equation}
where the average $\left<\cdots\right>$ is taken with respect to the
unperturbed system, and $\Delta S = S[\theta+\Delta\theta]-S[\theta]$.
Going back to the flux line representation \Eq{eq:vortexlinerepr2} we
have
\begin{eqnarray}
  \Delta S \nonumber &=& i \sum_\br \Delta\theta_\br (\nabla\cdot\bm_\br - qN_\br) \\
  &=& \nonumber - i \sum_\br \bm_\br \cdot \nabla \Delta\theta_\br  + \Delta\theta_\br qN_\br \\
  &=& - \f{2\pi i k}{q} \sum_{\br \in \mathcal{S}} \bm \cdot \hat\bn_\br
    - i \sum_\br \Delta\theta_\br qN_\br
\end{eqnarray}
where $\mathcal{S}$ is the surface at which $\Delta\theta$ is
discontinuous and $\hat\bn_\br$ is the surface normal at lattice site
$\br$.
Because $\Delta\theta_\br$ is an integer multiple of $2\pi/q$ only the
first term on the right hand side contribute when exponentiated.  The
resulting formula is more transparent when written in the continuum,
\begin{equation}
 \av{ e^{-\Delta S} } = 
 \av{ e^{\f{2\pi i k}{q} \int_\mathcal{S} \bm \cdot d\mathbf{S}} } =
 \av{ e^{\f{2\pi i k}{q} M_\nu } }.
\end{equation}
This is nothing but the large Wilson loop $\tilde{W}$ for a test
particle with \emph{fractional} charge $k$.
Thus the domain wall energy of the dual model is related to $\tilde W$
by
\begin{equation}
  \Delta F = - \ln \tilde{W}.
\end{equation}
In the (dual) ordered phase this will be proportional to the surface
area, $\Delta F \sim L^{d-1} = L^2$, consistent with an area law for
the Wilson loop $\tilde{W} \sim \exp({-c L^2})$.  In the
disordered phase, creating a domain wall costs no energy and $\Delta F
\to 0$ with system size.  Precisely at a critical point $\Delta F \to
$ a universal constant, independent of system size.  For a finite
sized system we expect the finite size scaling relation
\begin{equation}					\label{eq:scaling}
  \Delta F(\delta,L) = f(\delta L^{1/\nu})
\end{equation}
to hold, where $\delta = (K^{-1} - K_c^{-1})/K_c^{-1}$ is the tuning
parameter of the transition and $\nu$ the correlation length exponent.

From simulation data the critical point can thus be located by
plotting $\Delta F$ (or equivalently $\tilde{W}$) vs the tuning
parameter for different system sizes and observing the point where the
different curves cross.  We will show examples of this below in
\Fig{W}.

\subsection{Topological order}			\label{sec:topo}

One characteristic feature of systems displaying topological order is
a ground state degeneracy which depends on the topology of the space
on which the model is defined.  A further prediction is that for
finite sized systems the exact degeneracy will be lifted due to
tunneling events, leading to a ground state splitting $\Delta \sim
\exp{\left( - L / \xi \right)}$ \cite{wen91}.  In
Ref.~ \onlinecite{ves1} it was shown that there is a close relationship
between the lifting of the ground state degeneracy and the large
Wilson loop $\tilde W$ discussed above.  Here we give more details on
this relation and discuss the generalization to topologies other than
the torus.

\begin{figure}
\includegraphics[angle=-90,width=5cm]{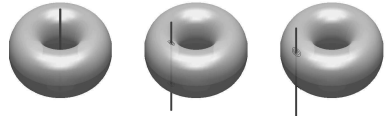}
\caption{\label{fig:munkar}
  A flux quantum tunnels out from the hole of a torus.
}
\end{figure}

To simplify the discussion we consider first the case $q=2$ on a
torus, with $\tilde W$ in the space-time plane.
In order to study ground state properties of the system we have to
consider anisotropic systems with size $L \times L \times \beta$ and
take the limit where the length of the time dimension $\beta$ goes to
infinity.
In the infinite system size limit the degenerate ground states may be
characterized by having an even or odd number of flux quanta $\Phi_0$
through the holes of the torus, giving a degeneracy of $2^2 = 4$ since
a torus has two holes~\cite{hos2004}.
The general case of arbitrary $q$ on a manifold with genus $g$ (i.e.,
which has $g$ handles and $2g$ holes) naturally leads to $q^{2g}$
degenerate ground states.
For finite system size the degeneracy will be lifted by vortex
tunneling processes, in which a vortex tunnels around a nontrivial
path on the torus as illustrated in \Fig{fig:munkar}.  In the
deconfined phase such processes are exponentially suppressed and the
system is topologically ordered.  In the confinement phase, however,
the vortices condense, and the tunneling is no longer negligible even
in the thermodynamic limit.
Let us concentrate on just one of the two holes of a torus for now.
We will denote the quantum mechanical states with an even or odd
number of flux quanta through the hole by $\ket{0}$ and $\ket{1}$,
respectively.

The space-time configurations of vortex lines and monopoles that
contribute to the partition function splits into different topological
sectors, depending on whether the winding number $M_\nu$ defined in
\Ref{eq:winding} is even or odd, i.e., $Z = Z_\mathrm{even} +
Z_\mathrm{odd}$.
Our Wilson loop $\tilde W$ (for $q=2$) can be written in terms of
these as
\begin{equation}					\label{eq:wilson-2}
  \tilde{W} = \av{e^{i \pi M_\nu}} = 
  \f{Z_\mathrm{even} - Z_\mathrm{odd}}{Z_\mathrm{even} + Z_\mathrm{odd}}.
\end{equation}
Quantum mechanically the configurations in the even sector contribute
to the diagonal matrix elements
\begin{equation}
  Z_\mathrm{even} = 
\bra{0} e^{-\beta \hat{H}}\ket{0}
  =
\bra{1} e^{-\beta \hat{H}}\ket{1},
\end{equation}
while the odd configurations contribute to tunneling matrix elements
\begin{equation}
  Z_\mathrm{odd} = 
\bra{0} e^{-\beta \hat{H}}\ket{1}
  =
\bra{1} e^{-\beta \hat{H}}\ket{0}.
\end{equation}
The tunneling lifts the degeneracy and mixes the different flux states.
%
% Tunneling Hamiltonian:
The Hamiltonian restricted to the (almost) degenerate ground state
Hilbert space is
\begin{equation}
  \hat{H} = \left( 
  \begin{array}{cc}
    E_0 & -\Delta \\
    -\Delta & E_0 \\
  \end{array}
  \right)
\end{equation}
which has energy eigenvalues $E = E_0 \mp \Delta$ and eigenstates
$\ket{\pm} = \left( \ket{0} \pm \ket{1} \right) / \sqrt{2}$.
The ground state splitting $\Delta$ can be related to
$\tilde W$ by the following equality, valid for $\beta \to \infty$:
\begin{equation}
  e^{-\beta \hat H} =
  e^{-\beta E_0} \left(
  \begin{array}{cc}
    \cosh \beta\Delta & 
    \sinh \beta\Delta \\ 
    \sinh \beta\Delta & 
    \cosh \beta\Delta \\
  \end{array}
  \right)
  =
  \left(
  \begin{array}{cc}
    Z_\mathrm{even} & Z_\mathrm{odd} \\
    Z_\mathrm{odd}  & Z_\mathrm{even} \\
  \end{array}
  \right).
\end{equation}
Using \Eq{eq:wilson-2} we have
\begin{equation}					\label{eq:W-Delta}
  \tilde{W} = e^{- 2 \beta \Delta },
\end{equation}
as found in Ref.~\onlinecite{ves1}.
In the deconfined phase $\tilde{W} = 1$ in the thermodynamic limit and
we must have $\Delta = 0$ for $\beta \to \infty$, i.e., an exact
degeneracy.  For finite $L$ the splitting can be evaluated directly
deep in the deconfinement phase, where the flux lines cost much
energy and are very dilute.  The dominating configurations which have
to be taken into account then are those with $n$ straight tunneling
trajectories across the system with action $S \approx \sigma L n$.
This gives
\begin{equation}
  Z \approx \sum_n \f{1}{n!}\left(\beta L\right)^n e^{-\sigma L n}
  = e^{\beta L e^{-\sigma L}},
\end{equation}
and
\begin{equation}
  \tilde W Z \approx \sum_n \f{(-1)^n}{n!}\left(\beta L\right)^n
  e^{-\sigma L n} = e^{-\beta L e^{-\sigma L}},
\end{equation}
where $\sigma$ is the line tension and there are $\beta L$ places to
put the tunneling trajectory.
In this limit $\tilde W$ will thus be given by $\tilde W \approx \exp{
  \left( - 2 \beta L e^{- \sigma L} \right)}$. 
Precisely such an exponential form of the ground state splitting has
been argued for by Wen~\cite{Wen90}.
Approaching the phase transition the flux line will no longer be
straight and interaction effects between different flux lines must be
taken into account, making it much more difficult to estimate the
ground state splitting.  However, matching to the scaling form
\Eq{eq:scaling} for $\tilde W$ suggests that $\tilde{W} = \exp{ \left(
- 2 \beta L/\xi^2 e^{- \sigma L /\xi} \right) }$, so that
\begin{equation}
\Delta = L/\xi^2 e^{- \sigma L /\xi },
\label{Deltascal}
\end{equation}
where $\xi \sim |\delta|^{-\nu}$ is the correlation length.  This
should hold for $L \gtrsim \xi$ on the deconfined side of the
transition.  Indeed, we will verify in the next sections that this
form is in agreement with the data from numerical simulations.
In the confinement phase an area law $\tilde{W} \sim e^{-c\beta
L}$ for the Wilson loop translates into a splitting $\Delta \sim L$
that grows with system size.

The arguments given above naturally generalize to manifolds of
arbitrary genus $g$ and charge $q$.  There are $2g$ nontrivial loops
around which vortex tunneling can take place.  Each such tunneling
event changes the flux through the corresponding hole by one flux
quantum $\Phi_0$.  In the infinite system size limit the $q^{2g}$ flux
states $\ket {\vec f} = \ket {f_1} \ket {f_2} \ldots \ket {f_{2g}}$,
where $f_i \in \mathbb{Z}_q$ is the number of flux quanta through hole
$i$ (mod $q$), are degenerate ground states.  The tunneling lifts the
degeneracy leading to a splitting
\begin{equation}					\label{eq:gensplit}
  \Delta E_{\vec k} = - \f{1}{\beta} \ln \tilde W_{\vec k}
\end{equation}
from the ground state.
Here we generalize \Eq{eq:W} to
\begin{equation}
\tilde W_{\vec k} = \left< \exp \left( \f{2\pi i}{q} \sum_{i=1}^{2g}
k_i M_i \right) \right>,
\end{equation}
where $M_i$ measures the flux quanta (mod $q$) tunneling through the
space-time surface surrounding hole $i$, and $k_i \in \mathbb{Z}_q$.
The corresponding eigenstates are given in the appendix in
\Eq{eq:eigenstates} and the ground state is the product of the
symmetric combinations of all the flux states
\begin{equation}
 \ket G = \prod_{i=1}^{2g} \left( \f{1}{\sqrt{q}}
 \sum_{f_i=0}^{q-1} \ket f_i \right).
\end{equation}
Away from the transition one may, to a very good approximation,
neglect higher order tunneling processes and only consider those which
tunnel single flux quanta.  Denoting those tunneling matrix elements
by $T_i$ the energy levels become $E_{\vec k} = E_0 - \sum_i T_i
2\cos\left( 2\pi k_i /q \right)$, i.e., the contributions from
individual tunneling processes are just added together.

\subsection{Vortex gap}

An important property of the ordered phase is that vortex excitations
are gapped.  This gap can be related to the large Wilson loop oriented
normal to the time direction $\tilde{W}_\tau$.  Returning to the $q=2$
case the vortex gap is
\begin{equation}					\label{Ev}
  E_v= - \lim_{\beta \to \infty} \frac{1}{\beta} \ln \frac{
    Z_\mathrm{odd}}{ Z_\mathrm{even}} = - \lim_{\beta \to \infty}
    \frac{1}{\beta} \ln \frac{1-\tilde{W}_\tau}{1+\tilde{W}_\tau},
\end{equation}
where $Z_\mathrm{odd}$ and $Z_\mathrm{even}$ now refer to the
partition functions with an odd and even number of vortices present.
A scaling argument analogous to \Eq{Deltascal} gives
\begin{equation}					\label{eq:Ev-scaling}
  \xi E_v= \sigma + \frac{2 \xi}{\beta}\ln \frac{L}{\xi},
\end{equation}
on the deconfined side of the transition in the low temperature limit
$\beta \to \infty$ and $\xi \lesssim L$.

\begin{figure*}
\includegraphics[width=7.2cm]{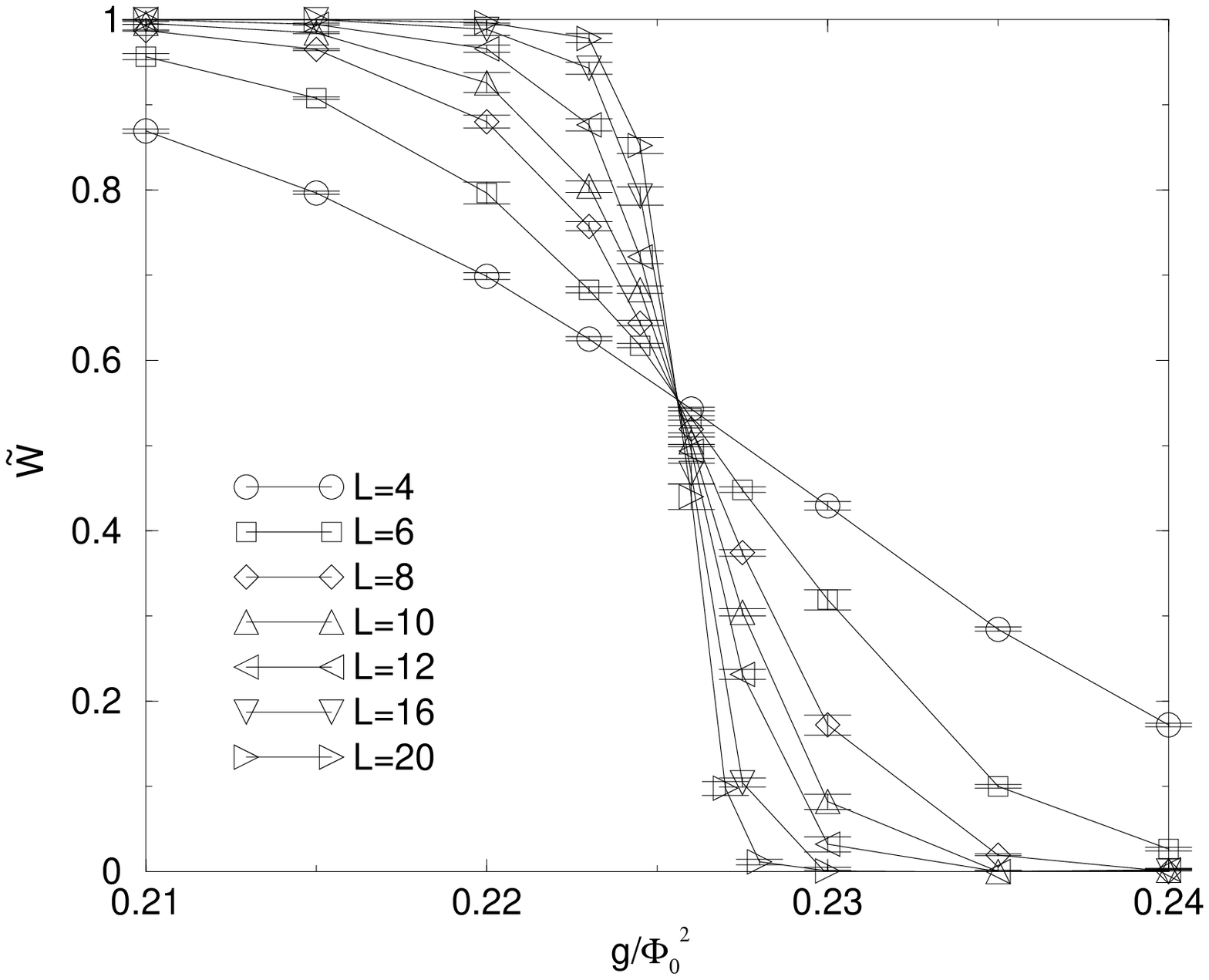}\includegraphics[width=7.2cm]{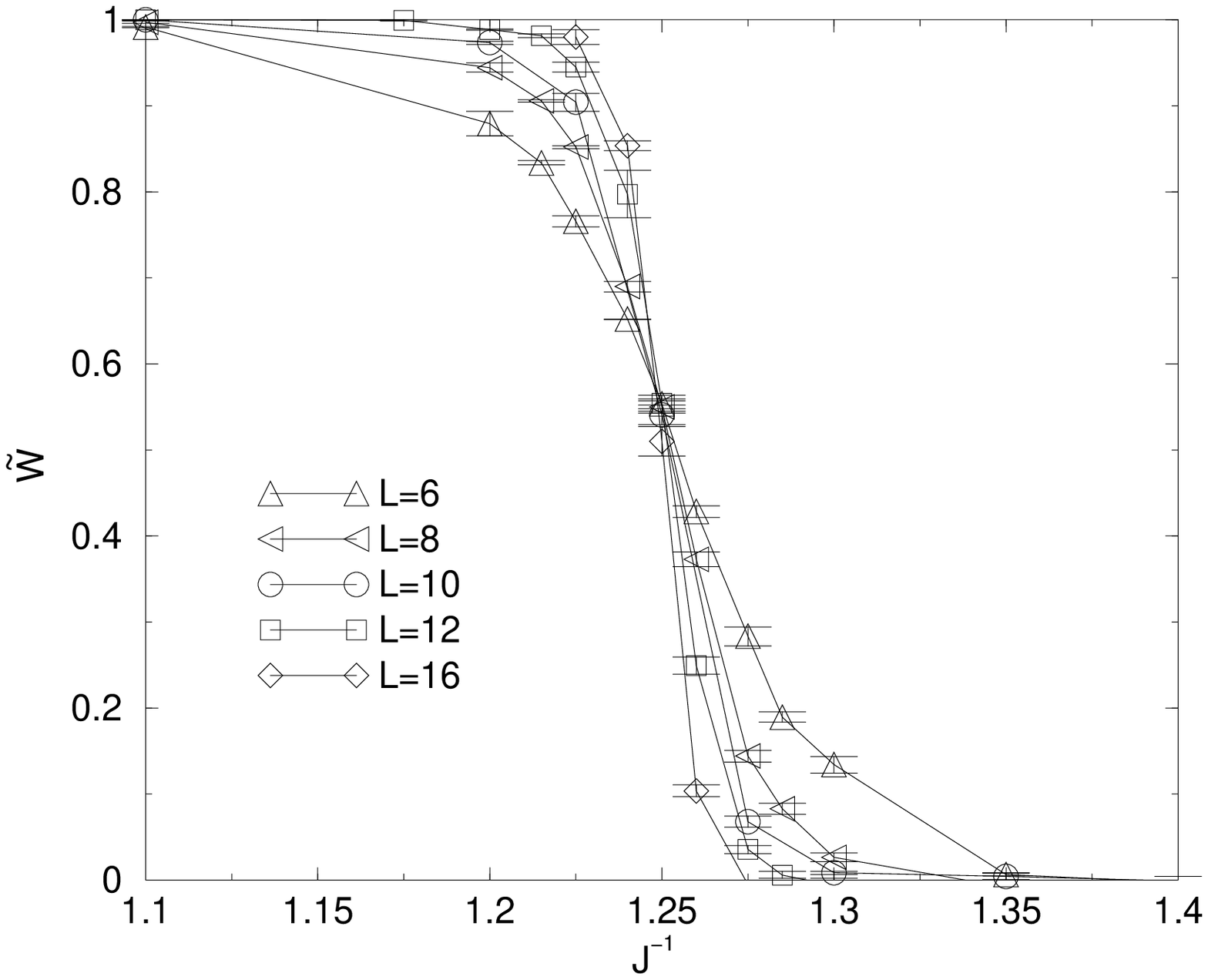}
\includegraphics[width=7.2cm]{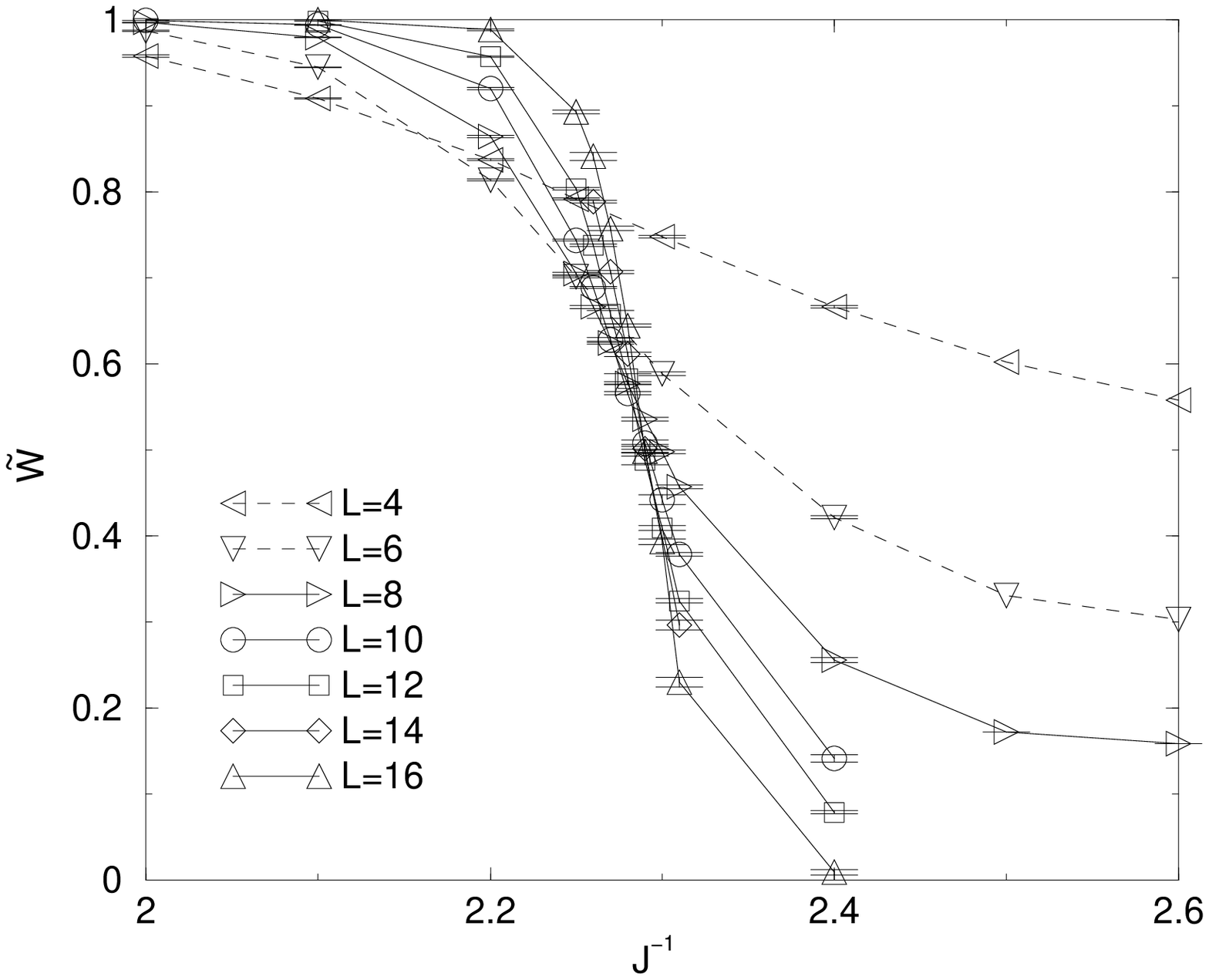}\includegraphics[width=7.2cm]{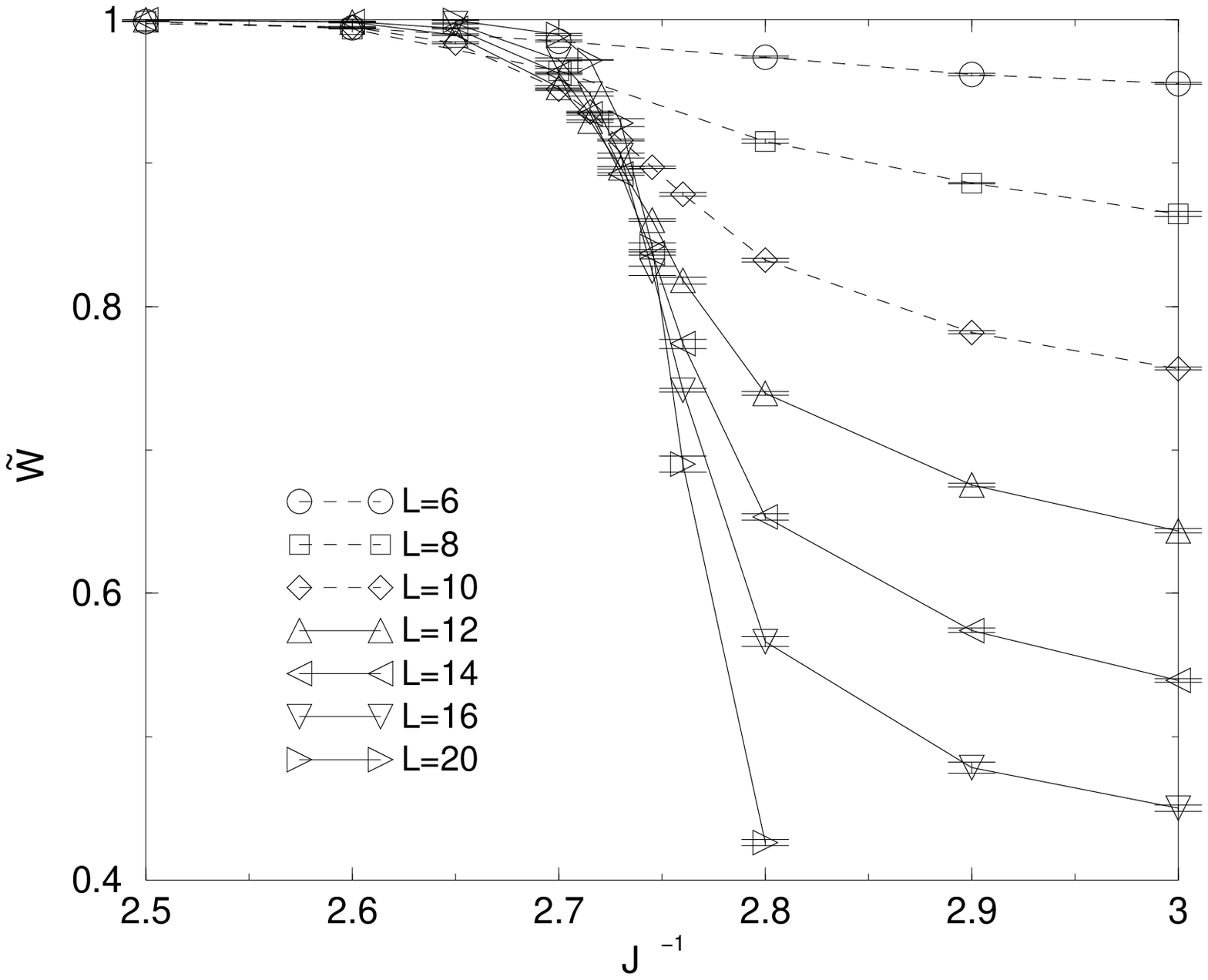}
\caption{
\label{W}
The large Wilson loop order parameter $\tilde{W}$ as function of
coupling constant, for different values of the screening length
$\lambda$.  Upper left: $\tilde W$ \emph{vs} $g/\Phi_0^2$ for
$\lambda=0$.  Upper right, lower left, and lower right: $\tilde W$
\emph{vs} $J^{-1}$ for $\lambda=0.5$, $1$, $2$, respectively.  The
critical point is where the curves for different system sizes cross.
Note that only system sizes much larger than $\lambda$ scale well
(solid lines), smaller sizes (dashed lines) suffers from finite size
corrections.}
\end{figure*}

\section{Monte Carlo Algorithm}				\label{sec:MC}

We use Monte Carlo simulations to obtain quantitative results for the
model and the nonlocal order parameter $\tilde W$.  The simulations
are performed in the loop-gas representation, \Eq{eq:vortexlinerepr},
of flux lines and monopoles, and we focus on the cases $q=2$ and $q=3$.
We use simple cubic lattices with periodic boundary conditions for the
vortices and monopoles, corresponding to fluctuating twist boundary
conditions in the phase representation~\cite{ves1}.

In the simulation we must update both the flux lines and the
monopoles.  In order to calculate $\tilde W$ it is necessary to
include global moves that change the winding numbers $M_\nu$ and thus
the topology of the vortex configuration.  In a conventional
Metropolis algorithm the acceptance ratio for such moves becomes
exponentially small with increased lattice size.  To overcome this we
use a worm cluster Monte Carlo algorithm \cite{worm,worms2}.  This
algorithm is described for on-site interactions in Ref.~
\onlinecite{worm}, but is straightforward to adapt to longer range
interactions.

To construct a worm one starts at a random lattice site $s_1$ in
space-time.  A vortex segment is then created in direction $\sigma$ to
a new lattice site $s_2$. The direction is chosen with the probability
\begin{equation}
  P_\sigma \sim A_\sigma=\min(1,\exp(- \Delta S_\sigma)),
\end{equation}
where $\Delta S_\sigma$ is the change in the action $S$ for creating a
vortex segment in direction $\sigma$. A new direction and lattice site
are then chosen from $s_2$. These steps are repeated until the worm of
vortex segments form a closed loop. To ensure detailed balance, the
final worm is accepted with the probability $P=\min[1,N ({\rm no
worm})/N({\rm worm})]$, where $N=\sum_\sigma A_\sigma$ at the initial
site, before and after the creation of the loop.  Occasionally the
loop constructed this way winds around the system, leading to the
desired changes in the winding number $M_\nu$.

The model allows magnetic monopoles in addition to the flux lines.
The monopoles are updated using a standard Metropolis scheme, with
trial moves consisting of the insertion of a nearest neighbor pair of
oppositely charged monopoles connected by $q$ flux quanta on the link
between them.  These moves are then accepted with probability
$\min(1,\exp(-\Delta S))$.  The creation of a plus monopole on top of
a minus monopole (or \emph{vice versa}) leads to their annihilation so
that the procedure described includes creation, destruction, as well
as motion of monopoles.

It is straightforward to show that the algorithm described above,
combining both the worm updates of the flux lines and the Metropolis
monopole moves, fulfills both ergodicity and detailed balance, and
respects the constraint $\nabla \cdot \bm = qN$.
The Wilson loop order parameter $\tilde W_\mu$ in direction $\mu$ is
calculated in the simulations as $\langle \cos(2\pi M_\mu/q) \rangle$
where $M_\mu$ is the number of vortex lines crossing a plane,
perpendicular to direction $\mu$, of the system.

% Results:

\section{Results}				\label{sec:results}

\begin{figure}
\centerline{\includegraphics[width=7.8cm]{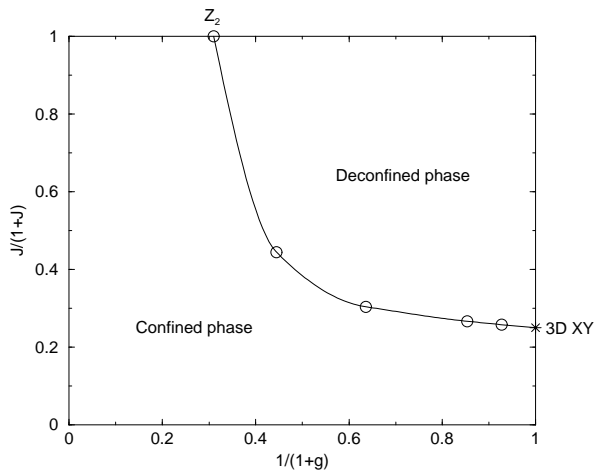}}
\caption{
\label{phase}
Phase diagram of the model with $q=2$.  Circles indicate the points
which have been simulated.  The parameters have been rescaled in order
to fit the phase diagram into the figure, so that $J\to \infty$
occur at the upper side and $g \to 0$ at the right side.  The
phase diagram connects the 3D $XY$ point at $g=0$ and the $Z_2$
transition at strong coupling $J\to\infty$.  The deconfined phase is
topologically ordered.  }
\end{figure}

For $q=2$ we have made simulations for constant
$\lambda=0,0.5,1.0,2.0$, varying $J$ (or $g$ in the $\lambda=0$
case) in each simulation.  The results for $\tilde{W}$ are shown in
\Fig{W}.  Below the transition $\tilde W$ goes to one with increasing
system size.  This is the deconfinement phase.  Above the transition,
in the confinement phase, it goes to zero.  Precisely at the
transition $\tilde W$ should be independent of system size according
to the scaling formula \Eq{eq:scaling}.
The figures clearly show how $\tilde{W}$ for system sizes much larger
than $\lambda$ cross at a given $J_c$.  However, it is also apparent
that for large values of $\lambda$ there are strong corrections to
scaling which influence the smaller sizes.  This is not surprising
since the interaction \Eq{eq:interaction} looks screened only at
distances much larger than $\lambda$.
It is important to have this in mind during the finite size scaling
analysis below.
The phase diagram determined from the simulations is shown in
\Fig{phase}.

\begin{figure}[b]
\centerline{\includegraphics[width=7.4cm]{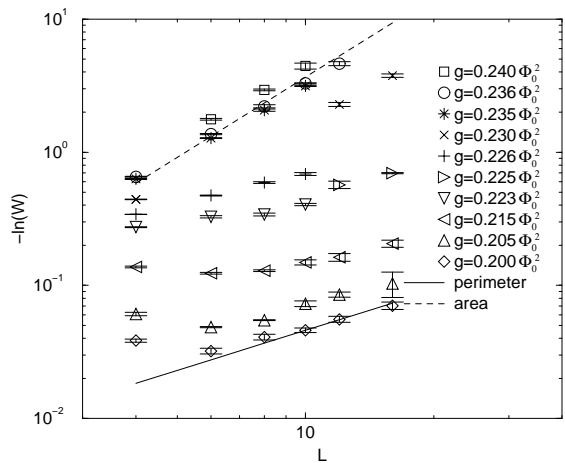}}
\caption{
\label{areaandperimeter}
Area law and perimeter law demonstrated for $\lambda \to 0$, for a
Wilson loop $W(\mathcal{C})$ which covers a quarter of a cross section
of the system and therefore has a finite perimeter and area.  Dashed
and solid lines are exact area and perimeter laws, inserted for
comparison.}
\end{figure}

As discussed $\tilde W$ differs from an ordinary Wilson loop in that
it does not follow a perimeter law in the deconfined phase.
In order to demonstrate both area and perimeter law, we use a Wilson
loop that covers one quarter of the system.  This configuration is
less suitable to locate $J_c$, but demonstrates the area and perimeter
laws nicely in \Fig{areaandperimeter}.

\begin{figure*}
\includegraphics[width=7.5cm]{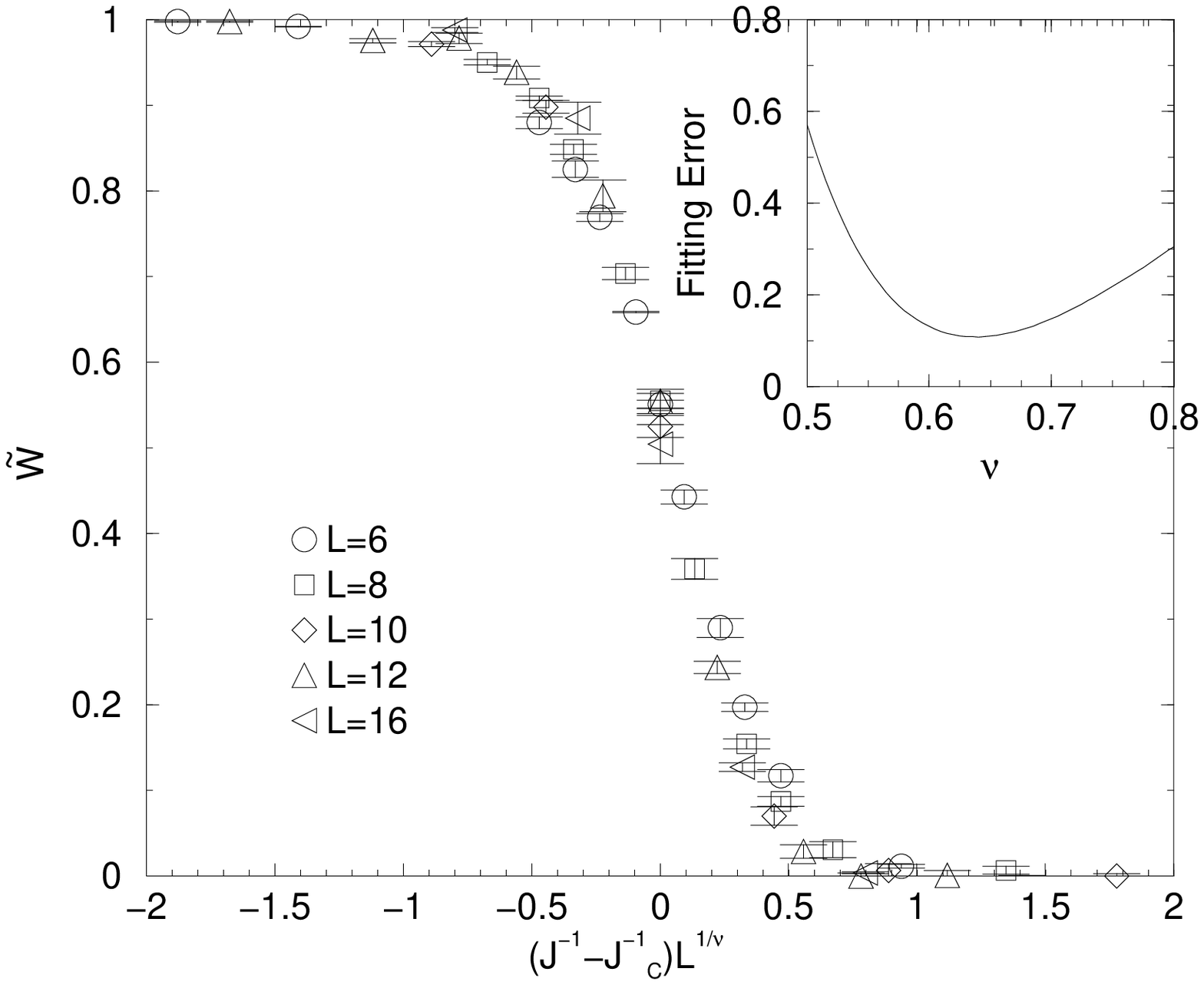}\includegraphics[width=7.5cm]{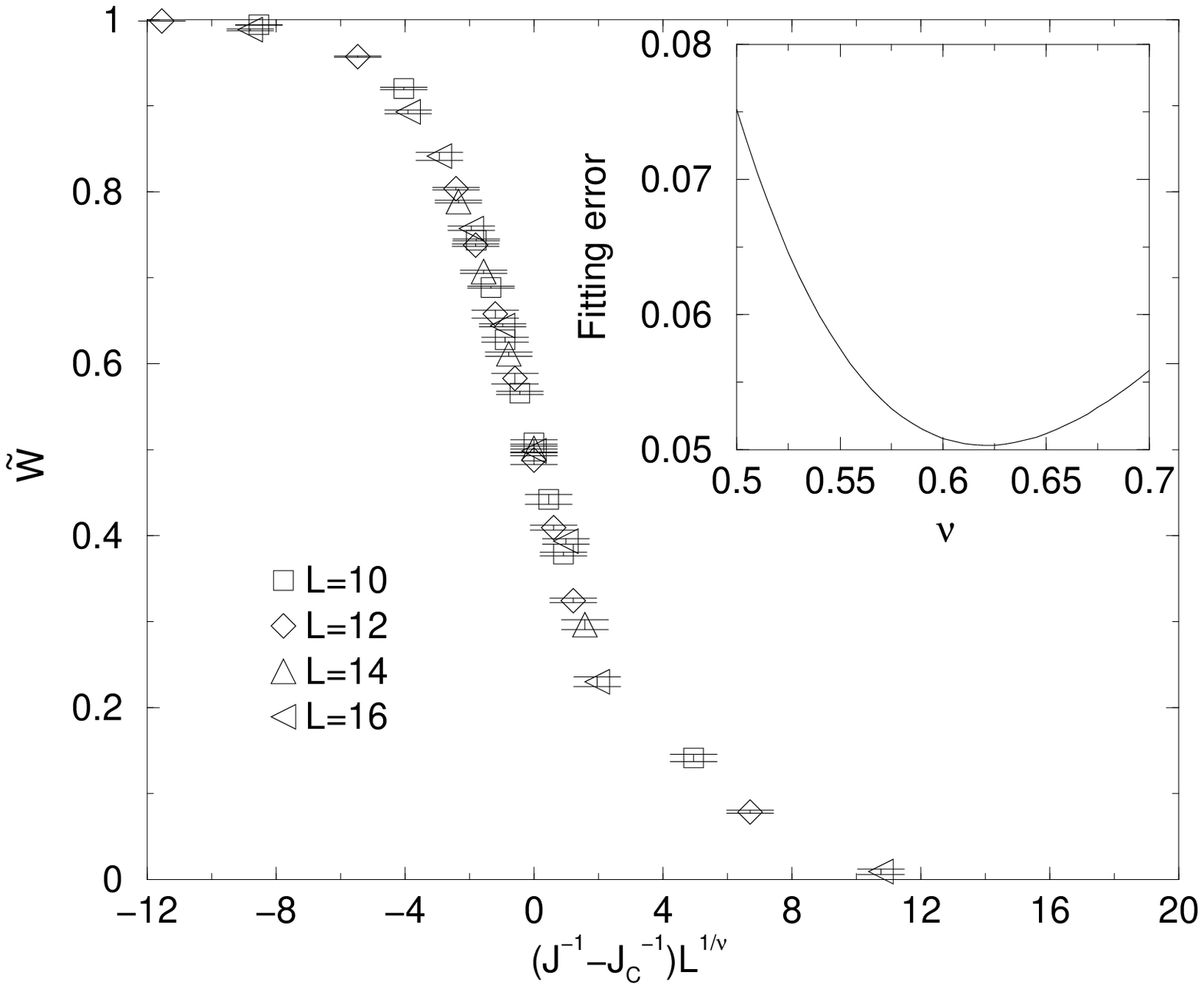}
\caption{
\label{FSS05}
\label{FSS1}
Finite size scaling plot of the $\tilde{W}$ order parameter for
$\lambda=0.5$ (left) with $\nu=0.64$ and $\lambda=1$ (right) with
$\nu=0.62$.  $J_c^{-1}$ is determined from \Fig{W} to 1.25 and 2.29,
respectively.  The inset shows fitting error as a function of $\nu$.}
\end{figure*}

\begin{figure}[b]
\centerline{\includegraphics[width=7.5cm]{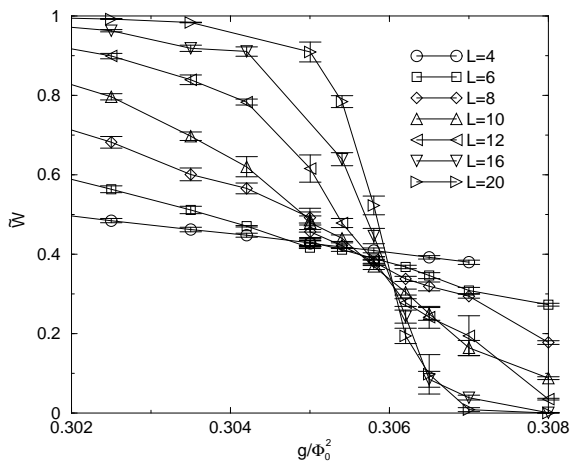}}
\caption{
\label{q3}
$\tilde{W}$ vs $g/\Phi_0^2$ for a model with $q=3$ and $\lambda=0$. Note
that this model has a first order transition and no scaling behavior
is expected.}
\end{figure}

We also simulate the model with $q=3$ to show that $\tilde W$ works
also for larger values of $q$.  The result for $\tilde W$ is shown in
\Fig{q3} for $\lambda=0$.  For $q=3$ the model has a first order
transition when $\lambda = 0$~\cite{smiseth}, and hence we expect no
scaling of $\tilde{W}$.  This is consistent with our MC data, since we
find no good crossing.  Instead the transition becomes sharper with
increasing system size and it is conceivable that a discontinuity
develops in the thermodynamic limit.

We now turn to the finite size scaling relation given in
\Eq{eq:scaling}, and the correlation length exponent $\nu$, for $q=2$.
With an appropriate choice of $\nu$ the data for different sizes
should collapse onto a single scaling function.  This is achieved by
minimizing
\begin{equation}					\label{minimi}
\sum_L\int
\left[\tilde{W}_L(x) - \overline{\tilde{W}}(x)
  \right]^2 dx,
\end{equation}
where $x=(J^{-1} - J_c^{-1}) L^{1/\nu}$ and $\overline{\tilde{W}}$ is
the mean value of $\tilde{W}_L$.  We calculate $\nu$ and $J_c$ by
making a multi parameter minimization of \Eq{minimi}.  Figure
\ref{FSS05} shows two of the resulting data collapses.  In doing this
minimization it is important to make use only of system sizes large
enough that corrections to scaling are negligible, hence we include
only those sizes for which well defined crossings were found above.
We get $\nu = 0.633 \pm 0.007$, $0.64 \pm 0.015$, $0.62 \pm 0.03$,
$0.6 \pm 0.15$, for $\lambda = 0, 0.5, 1, 2$, respectively.  Within
error-bars the data for all values of $\lambda$ that we have tested
are consistent with the expected $\nu \approx 0.63$ of the 3D Ising
model, which is the dual of the $Z_2$ gauge theory, i.e., the limiting
case obtained for $\lambda \to 0$.
This is in sharp contrast to the continuously varying exponents found
by Sudb{\o} {\em et al.}\cite{sudbo} and Smiseth {\em et al.}
\cite{smiseth} Their results were obtained from studies of the third
moment of the action and showed substantial deviations from the 3D
Ising value for $\lambda \gtrsim 0.5$.  The parameter values used in
these papers overlap to large extent with the present ones, although
there is a slight shift in the location of the phase diagram resulting
from our use of the Villain model [\Eq{eq:Villain}, or equivalently
\Ref{eq:vortexlinerepr}] instead of the cosine version \Eq{eq:action}.
This is not expected to change the critical exponents, however.

\begin{figure}[b]
\centerline{\includegraphics[width=7.5cm]{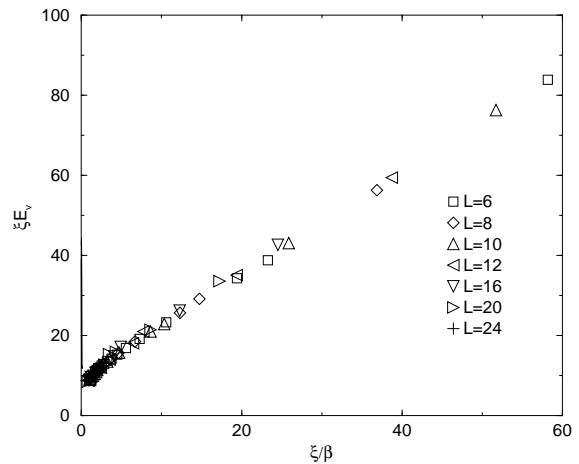}}
\caption{\label{vortexgap}
The energy gap to vortex excitations as a
function of temperature in the deconfined phase, for $q=2$ and $\lambda=0$.}
\end{figure}

Finally, we may also use $\tilde W$ to calculate two quantities which
are characteristic of topological order, namely the gap to vortex
excitations $E_v$, and the splitting $\Delta$ of the nearly degenerate
ground states on a torus.  Figure~\ref{vortexgap} shows $\xi E_v$
calculated using \Eq{Ev} as a function of $\xi/\beta$, where $\xi \sim
|\delta|^{-\nu}$.  The data nicely follows the scaling relation
proposed in \Eq{eq:Ev-scaling} in the deconfined phase, with $\sigma
\approx 10$.

\begin{figure}
\includegraphics[width=\linewidth]{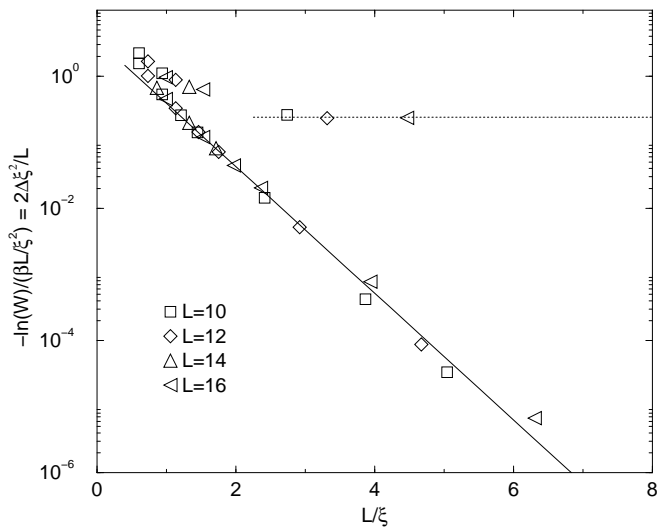}
\caption{
  \label{GSsplittwist}
  The scaling combination $2\Delta \xi^2/L$ plotted against $L/\xi$
  for $q=2$, $\lambda = 1$.  In the topologically ordered
  deconfinement phase the data follows the exponential dependence
  given in \Eq{Deltascal}, as indicated by the solid line.  The dashed
  line indicates the area law obeyed in the confinement phase. }
\end{figure}

The relation between $\tilde W$ and $\Delta$ in \Eq{eq:W-Delta} allows
the ground state splitting to be explicitly calculated using the Monte
Carlo data.  An example of this is shown in \Fig{GSsplittwist}, where
the exponential dependence on system size obtained in \Eq{Deltascal}
is clearly seen in the deconfinement phase (solid line), while an area
law (dashed line) is found in the confinement phase.  For further
examples and discussion of this aspect see Ref.~ \onlinecite{ves1}.

\section{Summary}				\label{sec:summary}

In summary, we have studied a compact $U(1)$ gauge theory coupled to
bosonic matter with gauge charge $q$ using duality arguments and Monte
Carlo simulations.  The confinement--deconfinement transition is
analyzed using a nonlocal order parameter $\tilde W$, which is related
to large Wilson loops with the loop covering a whole cross section of
the system, and which directly probes the confining properties of the
theory.  The model can be mapped to a dual model with a global
$q$-state clock symmetry, and $\tilde W$ corresponds to the domain
wall energy of that model, which suggests a scaling behavior according
to \Eq{eq:scaling}.
We confirm this using Monte Carlo simulations and study the details of
the deconfinement transition over a range of parameters.  Our results
for the critical exponent $\nu$ support a transition in the 3D Ising
universality class for all values of $\lambda$ tested.  In particular,
we do not see any trace of continuously varying critical exponents
found in Refs.~ \onlinecite{smiseth,sudbo} using a different order
parameter.  There are, however, strong crossover effects in the
vicinity of the 3D $XY$ transition at $\lambda\to\infty$, which make
an accurate determination of the exponents tricky in this limit.
The deconfined phase is topologically ordered with gapped vortex
excitations, with a gap $E_v \sim \xi^{-1}$, which can be related to
$\tilde W$ for a spatial loop.  The ground state degeneracy
characteristic of topological order is lifted in finite sized systems
with a gap $2\Delta$ which is simply related to $\tilde W$ for a
space-time loop~\cite{ves1}.
This establishes a firm relationship between topological order and
deconfinement and allows explicit calculations of the ground state
energy splitting in the topologically ordered phase of the theory.  We
expect the methods developed here to be useful for quantitative
studies of topological order in other systems as well.

\begin{acknowledgments}
We thank T.\ Hans Hanson, and Asle Sudb{\o} for valuable
discussions. Support from the Swedish Research Council (VR) and the
G{\"o}ran Gustafsson foundation is gratefully acknowledged.
\end{acknowledgments}

% \vspace{10mm}

\appendix*

\section{Ground state splitting for general $g$ and $q$}

In this appendix we give a detailed derivation of the relation between
the ground state splitting and $\tilde W$ for models with gauge charge
$q$ on manifolds with arbitrary genus $g$.  The configurations
contributing to the partition function can be split into separate
topological sectors depending on the flux $\Phi_0 m_i$ passing through
a space-time cross section surrounding hole $i$.  Let $Z_{\vec m} =
\bra{\vec f} e^{-\beta H} \ket{\vec f + \vec m}$ denote the partition
function restricted to the sector labeled by $\vec{m}$.  Then $\tilde
W_{\vec k} = \sum_{\vec m} Z_{\vec m} e^{2\pi i \vec k \cdot \vec m /
q} / \sum_{\vec m} Z_{\vec m}$.  The Hamiltonian in the ground state
subspace can be written as
\begin{equation}
  \cH = \sum_{\vec m} h_{\vec m} \sum_{\vec f} \ket {\vec f} \bra
      {\vec f + \vec m} + \ket {\vec f + \vec m} \bra {\vec f},
\end{equation}
where $h_{\vec 0} = E_0/2$ and $h_{\vec m} = T_{\vec m}$ ($\vec m \ne
\vec 0$) are tunneling amplitudes.  Because of translation invariance
in the index $\vec f$ the eigenstates are
\begin{equation}					\label{eq:eigenstates}
  \ket {\tilde k} = \f{1}{\sqrt{q^{2g}}} \sum_{\vec f} e^{2\pi i \vec k
  \cdot \vec f /q} \ket {\vec f}
\end{equation}
with corresponding eigenvalues
\begin{equation}
  E_{\vec k} = \sum_{\vec m} h_{\vec m} 2\cos\left(2\pi \vec k \cdot
  \vec m / q \right).
\end{equation}
To relate the energies to the $\tilde W_{\vec k}$ that are easily
calculated in a simulation we use
\begin{eqnarray}
  e^{-\beta E_{\vec k}} &=& \bra{\tilde k} e^{-\beta \cH} \ket{\tilde
    k} = \\ \nonumber
  && \f{1}{q^{2g}} \sum_{\vec f, \vec f'} \bra{\vec f'} e^{-\beta
  \cH} \ket{\vec f} e^{2\pi i \vec k \cdot(\vec f - \vec f')/q} = \\ \nonumber
  && \sum_{\vec m} Z_{\vec m} e^{2\pi i \vec k \cdot \vec m /q} = \tilde
  W_{\vec k } \sum_{\vec m} Z_{\vec m},
\end{eqnarray}
from which \Eq{eq:gensplit} follows.

\bibliography{compact}

\begin{thebibliography}{35}
\expandafter\ifx\csname natexlab\endcsname\relax\def\natexlab#1{#1}\fi
\expandafter\ifx\csname bibnamefont\endcsname\relax
  \def\bibnamefont#1{#1}\fi
\expandafter\ifx\csname bibfnamefont\endcsname\relax
  \def\bibfnamefont#1{#1}\fi
\expandafter\ifx\csname citenamefont\endcsname\relax
  \def\citenamefont#1{#1}\fi
\expandafter\ifx\csname url\endcsname\relax
  \def\url#1{\texttt{#1}}\fi
\expandafter\ifx\csname urlprefix\endcsname\relax\def\urlprefix{URL }\fi
\providecommand{\bibinfo}[2]{#2}
\providecommand{\eprint}[2][]{\url{#2}}

\bibitem[{\citenamefont{Baskaran and Anderson}(1988)}]{baskaran:580}
\bibinfo{author}{\bibfnamefont{G.}~\bibnamefont{Baskaran}} \bibnamefont{and}
  \bibinfo{author}{\bibfnamefont{P.~W.} \bibnamefont{Anderson}},
  \bibinfo{journal}{\prb} \textbf{\bibinfo{volume}{37}}, \bibinfo{pages}{R580}
  (\bibinfo{year}{1988}).

\bibitem[{\citenamefont{Ioffe and Larkin}(1989)}]{ioffe:8988}
\bibinfo{author}{\bibfnamefont{L.~B.} \bibnamefont{Ioffe}} \bibnamefont{and}
  \bibinfo{author}{\bibfnamefont{A.~I.} \bibnamefont{Larkin}},
  \bibinfo{journal}{\prb} \textbf{\bibinfo{volume}{39}}, \bibinfo{pages}{8988}
  (\bibinfo{year}{1989}).

\bibitem[{\citenamefont{Nagaosa and Lee}(1992)}]{nagaosa:966}
\bibinfo{author}{\bibfnamefont{N.}~\bibnamefont{Nagaosa}} \bibnamefont{and}
  \bibinfo{author}{\bibfnamefont{P.~A.} \bibnamefont{Lee}},
  \bibinfo{journal}{\prb} \textbf{\bibinfo{volume}{45}}, \bibinfo{pages}{966}
  (\bibinfo{year}{1992}).

\bibitem[{\citenamefont{Lee and Nagaosa}(1992)}]{lee:5621}
\bibinfo{author}{\bibfnamefont{P.~A.} \bibnamefont{Lee}} \bibnamefont{and}
  \bibinfo{author}{\bibfnamefont{N.}~\bibnamefont{Nagaosa}},
  \bibinfo{journal}{\prb} \textbf{\bibinfo{volume}{46}}, \bibinfo{pages}{5621}
  (\bibinfo{year}{1992}).

\bibitem[{\citenamefont{Nagaosa and Lee}(2000)}]{nagaosa-lee}
\bibinfo{author}{\bibfnamefont{N.}~\bibnamefont{Nagaosa}} \bibnamefont{and}
  \bibinfo{author}{\bibfnamefont{P.~A.} \bibnamefont{Lee}},
  \bibinfo{journal}{\prb} \textbf{\bibinfo{volume}{61}}, \bibinfo{pages}{9166}
  (\bibinfo{year}{2000}).

\bibitem[{\citenamefont{Read and Sachdev}(1991)}]{Read91}
\bibinfo{author}{\bibfnamefont{N.}~\bibnamefont{Read}} \bibnamefont{and}
  \bibinfo{author}{\bibfnamefont{S.}~\bibnamefont{Sachdev}},
  \bibinfo{journal}{\prl} \textbf{\bibinfo{volume}{66}}, \bibinfo{pages}{1773}
  (\bibinfo{year}{1991}).

\bibitem[{\citenamefont{Sachdev and Read}(1991)}]{Sachdev91}
\bibinfo{author}{\bibfnamefont{S.}~\bibnamefont{Sachdev}} \bibnamefont{and}
  \bibinfo{author}{\bibfnamefont{N.}~\bibnamefont{Read}},
  \bibinfo{journal}{Int. J. Mod. Phys. B} \textbf{\bibinfo{volume}{5}},
  \bibinfo{pages}{219} (\bibinfo{year}{1991}).

\bibitem[{\citenamefont{Sachdev}(2003)}]{sachdev03}
\bibinfo{author}{\bibfnamefont{S.}~\bibnamefont{Sachdev}},
  \bibinfo{journal}{\rmp} \textbf{\bibinfo{volume}{75}}, \bibinfo{eid}{913}
  (\bibinfo{year}{2003}).

\bibitem[{\citenamefont{Wen and Lee}(1996)}]{wen:503}
\bibinfo{author}{\bibfnamefont{X.-G.} \bibnamefont{Wen}} \bibnamefont{and}
  \bibinfo{author}{\bibfnamefont{P.~A.} \bibnamefont{Lee}},
  \bibinfo{journal}{\prl} \textbf{\bibinfo{volume}{76}}, \bibinfo{pages}{503}
  (\bibinfo{year}{1996}).

\bibitem[{\citenamefont{Senthil and Fisher}(2001)}]{senthil-fisher}
\bibinfo{author}{\bibfnamefont{T.}~\bibnamefont{Senthil}} \bibnamefont{and}
  \bibinfo{author}{\bibfnamefont{M.~P.~A.} \bibnamefont{Fisher}},
  \bibinfo{journal}{\prb} \textbf{\bibinfo{volume}{63}}, \bibinfo{eid}{134521}
  (\bibinfo{year}{2001}).

\bibitem[{\citenamefont{Mudry and Fradkin}(1994{\natexlab{a}})}]{mudry-fradkin}
\bibinfo{author}{\bibfnamefont{C.}~\bibnamefont{Mudry}} \bibnamefont{and}
  \bibinfo{author}{\bibfnamefont{E.}~\bibnamefont{Fradkin}},
  \bibinfo{journal}{\prb} \textbf{\bibinfo{volume}{49}}, \bibinfo{pages}{5200}
  (\bibinfo{year}{1994}{\natexlab{a}}).

\bibitem[{\citenamefont{Mudry and
  Fradkin}(1994{\natexlab{b}})}]{mudry-fradkin-2}
\bibinfo{author}{\bibfnamefont{C.}~\bibnamefont{Mudry}} \bibnamefont{and}
  \bibinfo{author}{\bibfnamefont{E.}~\bibnamefont{Fradkin}},
  \bibinfo{journal}{\prb} \textbf{\bibinfo{volume}{50}}, \bibinfo{pages}{11409}
  (\bibinfo{year}{1994}{\natexlab{b}}).

\bibitem[{\citenamefont{Lee}(2000)}]{DHLee00}
\bibinfo{author}{\bibfnamefont{D.-H.} \bibnamefont{Lee}},
  \bibinfo{journal}{Phys. Rev. Lett.} \textbf{\bibinfo{volume}{84}},
  \bibinfo{pages}{2694} (\bibinfo{year}{2000}).

\bibitem[{\citenamefont{Ichinose et~al.}(2001)\citenamefont{Ichinose, Matsui,
  and Onoda}}]{Ichinose2001}
\bibinfo{author}{\bibfnamefont{I.}~\bibnamefont{Ichinose}},
  \bibinfo{author}{\bibfnamefont{T.}~\bibnamefont{Matsui}}, \bibnamefont{and}
  \bibinfo{author}{\bibfnamefont{M.}~\bibnamefont{Onoda}},
  \bibinfo{journal}{\prb} \textbf{\bibinfo{volume}{64}}, \bibinfo{eid}{104516}
  (\bibinfo{year}{2001}).

\bibitem[{\citenamefont{Senthil and Motrunich}(2002)}]{senthil-montro}
\bibinfo{author}{\bibfnamefont{T.}~\bibnamefont{Senthil}} \bibnamefont{and}
  \bibinfo{author}{\bibfnamefont{O.}~\bibnamefont{Motrunich}},
  \bibinfo{journal}{\prb} \textbf{\bibinfo{volume}{66}}, \bibinfo{eid}{205104}
  (\bibinfo{year}{2002}).

\bibitem[{\citenamefont{Wen}(1991{\natexlab{a}})}]{wen91}
\bibinfo{author}{\bibfnamefont{X.-G.} \bibnamefont{Wen}},
  \bibinfo{journal}{Int. Journ. Mod. Phys. B} \textbf{\bibinfo{volume}{5}},
  \bibinfo{pages}{1641} (\bibinfo{year}{1991}{\natexlab{a}}).

\bibitem[{\citenamefont{Ioffe et~al.}(2002)\citenamefont{Ioffe, Feigel'man,
  Ioselevich, Ivanov, Troyer, and Blatter}}]{Feigleman-Ioffe}
\bibinfo{author}{\bibfnamefont{L.~B.} \bibnamefont{Ioffe}},
  \bibinfo{author}{\bibfnamefont{M.~V.} \bibnamefont{Feigel'man}},
  \bibinfo{author}{\bibfnamefont{A.}~\bibnamefont{Ioselevich}},
  \bibinfo{author}{\bibfnamefont{D.}~\bibnamefont{Ivanov}},
  \bibinfo{author}{\bibfnamefont{M.}~\bibnamefont{Troyer}}, \bibnamefont{and}
  \bibinfo{author}{\bibfnamefont{G.}~\bibnamefont{Blatter}},
  \bibinfo{journal}{Nature} \textbf{\bibinfo{volume}{415}},
  \bibinfo{pages}{503} (\bibinfo{year}{2002}).

\bibitem[{\citenamefont{Kitaev}(2003)}]{kitaev}
\bibinfo{author}{\bibfnamefont{A.~Y.} \bibnamefont{Kitaev}},
  \bibinfo{journal}{Ann. Phys.} \textbf{\bibinfo{volume}{303}},
  \bibinfo{pages}{2} (\bibinfo{year}{2003}).

\bibitem[{\citenamefont{Franz and Tesanovic}(2001)}]{tesanovic}
\bibinfo{author}{\bibfnamefont{M.}~\bibnamefont{Franz}} \bibnamefont{and}
  \bibinfo{author}{\bibfnamefont{Z.}~\bibnamefont{Tesanovic}},
  \bibinfo{journal}{\prl} \textbf{\bibinfo{volume}{87}},
  \bibinfo{pages}{257003} (\bibinfo{year}{2001}).

\bibitem[{\citenamefont{Herbut}(2002)}]{herbut}
\bibinfo{author}{\bibfnamefont{I.~F.} \bibnamefont{Herbut}},
  \bibinfo{journal}{\prb} \textbf{\bibinfo{volume}{66}},
  \bibinfo{pages}{094504} (\bibinfo{year}{2002}).

\bibitem[{\citenamefont{Herbut and Seradjeh}(2003)}]{herbut-seradjeh}
\bibinfo{author}{\bibfnamefont{I.~F.} \bibnamefont{Herbut}} \bibnamefont{and}
  \bibinfo{author}{\bibfnamefont{B.~H.} \bibnamefont{Seradjeh}},
  \bibinfo{journal}{Phys. Rev. Lett.} \textbf{\bibinfo{volume}{91}},
  \bibinfo{eid}{171601} (\bibinfo{year}{2003}).

\bibitem[{\citenamefont{Hermele et~al.}(2004)\citenamefont{Hermele, Senthil,
  Fisher, Lee, Nagaosa, and Wen}}]{hermele}
\bibinfo{author}{\bibfnamefont{M.}~\bibnamefont{Hermele}},
  \bibinfo{author}{\bibfnamefont{T.}~\bibnamefont{Senthil}},
  \bibinfo{author}{\bibfnamefont{M.~P.~A.} \bibnamefont{Fisher}},
  \bibinfo{author}{\bibfnamefont{P.~A.} \bibnamefont{Lee}},
  \bibinfo{author}{\bibfnamefont{N.}~\bibnamefont{Nagaosa}}, \bibnamefont{and}
  \bibinfo{author}{\bibfnamefont{X.-G.} \bibnamefont{Wen}},
  \bibinfo{journal}{\prb} \textbf{\bibinfo{volume}{70}}, \bibinfo{eid}{214437}
  (\bibinfo{year}{2004}).

\bibitem[{\citenamefont{Kleinert et~al.}(2002)\citenamefont{Kleinert, Nogueira,
  and Sudb{\o}}}]{kleinert-nogu-sudbo}
\bibinfo{author}{\bibfnamefont{H.}~\bibnamefont{Kleinert}},
  \bibinfo{author}{\bibfnamefont{F.~S.} \bibnamefont{Nogueira}},
  \bibnamefont{and} \bibinfo{author}{\bibfnamefont{A.}~\bibnamefont{Sudb{\o}}},
  \bibinfo{journal}{\prl} \textbf{\bibinfo{volume}{88}}, \bibinfo{eid}{232001}
  (\bibinfo{year}{2002}).

\bibitem[{\citenamefont{Polyakov}(1977)}]{poly77}
\bibinfo{author}{\bibfnamefont{A.~M.} \bibnamefont{Polyakov}},
  \bibinfo{journal}{Nucl. Phys.} \textbf{\bibinfo{volume}{120}},
  \bibinfo{pages}{429} (\bibinfo{year}{1977}).

\bibitem[{\citenamefont{Fradkin and Shenker}(1979)}]{frad79}
\bibinfo{author}{\bibfnamefont{E.}~\bibnamefont{Fradkin}} \bibnamefont{and}
  \bibinfo{author}{\bibfnamefont{S.~H.} \bibnamefont{Shenker}},
  \bibinfo{journal}{\prd} \textbf{\bibinfo{volume}{19}}, \bibinfo{pages}{3682}
  (\bibinfo{year}{1979}).

\bibitem[{\citenamefont{Wen}(1991{\natexlab{b}})}]{wen-spinon-pairing}
\bibinfo{author}{\bibfnamefont{X.~G.} \bibnamefont{Wen}},
  \bibinfo{journal}{\prb} \textbf{\bibinfo{volume}{44}}, \bibinfo{pages}{2664}
  (\bibinfo{year}{1991}{\natexlab{b}}).

\bibitem[{\citenamefont{Sudb{\o} et~al.}(2002)\citenamefont{Sudb{\o},
  Sm{\o}rgrav, Smiseth, Nogueira, and Hove}}]{sudbo}
\bibinfo{author}{\bibfnamefont{A.}~\bibnamefont{Sudb{\o}}},
  \bibinfo{author}{\bibfnamefont{E.}~\bibnamefont{Sm{\o}rgrav}},
  \bibinfo{author}{\bibfnamefont{J.}~\bibnamefont{Smiseth}},
  \bibinfo{author}{\bibfnamefont{F.~S.} \bibnamefont{Nogueira}},
  \bibnamefont{and} \bibinfo{author}{\bibfnamefont{J.}~\bibnamefont{Hove}},
  \bibinfo{journal}{\prl} \textbf{\bibinfo{volume}{89}}, \bibinfo{eid}{226403}
  (\bibinfo{year}{2002}).

\bibitem[{\citenamefont{Smiseth et~al.}(2003)\citenamefont{Smiseth,
  Sm{\o}rgrav, Nogueira, Hove, and Sudb{\o}}}]{smiseth}
\bibinfo{author}{\bibfnamefont{J.}~\bibnamefont{Smiseth}},
  \bibinfo{author}{\bibfnamefont{E.}~\bibnamefont{Sm{\o}rgrav}},
  \bibinfo{author}{\bibfnamefont{F.~S.} \bibnamefont{Nogueira}},
  \bibinfo{author}{\bibfnamefont{J.}~\bibnamefont{Hove}}, \bibnamefont{and}
  \bibinfo{author}{\bibfnamefont{A.}~\bibnamefont{Sudb{\o}}},
  \bibinfo{journal}{\prb} \textbf{\bibinfo{volume}{67}}, \bibinfo{eid}{205104}
  (\bibinfo{year}{2003}).

\bibitem[{\citenamefont{Vestergren et~al.}(2005)\citenamefont{Vestergren,
  Lidmar, and Hansson}}]{ves1}
\bibinfo{author}{\bibfnamefont{A.}~\bibnamefont{Vestergren}},
  \bibinfo{author}{\bibfnamefont{J.}~\bibnamefont{Lidmar}}, \bibnamefont{and}
  \bibinfo{author}{\bibfnamefont{T.~H.} \bibnamefont{Hansson}},
  \bibinfo{journal}{Europhys. Lett.} \textbf{\bibinfo{volume}{69}},
  \bibinfo{pages}{256} (\bibinfo{year}{2005}).

\bibitem[{\citenamefont{Einhorn and Savit}(1978)}]{ein78}
\bibinfo{author}{\bibfnamefont{M.~B.} \bibnamefont{Einhorn}} \bibnamefont{and}
  \bibinfo{author}{\bibfnamefont{R.}~\bibnamefont{Savit}},
  \bibinfo{journal}{\prd} \textbf{\bibinfo{volume}{17}}, \bibinfo{pages}{2583}
  (\bibinfo{year}{1978}).

\bibitem[{\citenamefont{Chernodub et~al.}(2005)\citenamefont{Chernodub,
  Feldmann, Ilgenfritz, and Schiller}}]{chernodub}
\bibinfo{author}{\bibfnamefont{M.~N.} \bibnamefont{Chernodub}},
  \bibinfo{author}{\bibfnamefont{R.}~\bibnamefont{Feldmann}},
  \bibinfo{author}{\bibfnamefont{E.-M.} \bibnamefont{Ilgenfritz}},
  \bibnamefont{and} \bibinfo{author}{\bibfnamefont{A.}~\bibnamefont{Schiller}},
  \bibinfo{journal}{Phys.\ Lett.\ B} \textbf{\bibinfo{volume}{605}},
  \bibinfo{pages}{161} (\bibinfo{year}{2005}), \eprint{arXiv:hep-lat/0502009}.

\bibitem[{\citenamefont{Hansson et~al.}(2004)\citenamefont{Hansson, Oganesyan,
  and Sondhi}}]{hos2004}
\bibinfo{author}{\bibfnamefont{T.~H.} \bibnamefont{Hansson}},
  \bibinfo{author}{\bibfnamefont{V.}~\bibnamefont{Oganesyan}},
  \bibnamefont{and} \bibinfo{author}{\bibfnamefont{S.~L.}
  \bibnamefont{Sondhi}}, \bibinfo{journal}{Ann. Phys.}
  \textbf{\bibinfo{volume}{313}}, \bibinfo{pages}{497} (\bibinfo{year}{2004}).

\bibitem[{\citenamefont{Wen and Niu}(1990)}]{Wen90}
\bibinfo{author}{\bibfnamefont{X.-G.} \bibnamefont{Wen}} \bibnamefont{and}
  \bibinfo{author}{\bibfnamefont{Q.}~\bibnamefont{Niu}},
  \bibinfo{journal}{\prb} \textbf{\bibinfo{volume}{41}}, \bibinfo{pages}{9377}
  (\bibinfo{year}{1990}).

\bibitem[{\citenamefont{Alet and S{\o}rensen}(2003)}]{worm}
\bibinfo{author}{\bibfnamefont{F.}~\bibnamefont{Alet}} \bibnamefont{and}
  \bibinfo{author}{\bibfnamefont{E.~S.} \bibnamefont{S{\o}rensen}},
  \bibinfo{journal}{\pre} \textbf{\bibinfo{volume}{67}},
  \bibinfo{pages}{015701(R)} (\bibinfo{year}{2003}).

\bibitem[{\citenamefont{Prokof'ev et~al.}(1998)\citenamefont{Prokof'ev,
  Svistunov, and Tupitsyn}}]{worms2}
\bibinfo{author}{\bibfnamefont{N.~V.} \bibnamefont{Prokof'ev}},
  \bibinfo{author}{\bibfnamefont{B.~V.} \bibnamefont{Svistunov}},
  \bibnamefont{and} \bibinfo{author}{\bibfnamefont{I.~S.}
  \bibnamefont{Tupitsyn}}, \bibinfo{journal}{Phys. Lett. A}
  \textbf{\bibinfo{volume}{238}}, \bibinfo{pages}{253} (\bibinfo{year}{1998}).

\end{thebibliography}

\end{document}